\documentclass[twocolumn,eqsecnum,aps]{revtex4}
\usepackage[dvips]{epsfig,color}
\usepackage{graphicx}
\usepackage{amsmath,amsfonts,amsbsy,amssymb}
\usepackage{tabularx}

\begin{document}

\title{Quantum Hall Liquid on a Noncommutative  Superplane}

\author{Kazuki Hasebe}
\affiliation{Yukawa Institute for Theoretical physics, Kyoto
 University,
 Kyoto 606-8502, Japan \\
Email: hasebe@yukawa.kyoto-u.ac.jp}

\begin{abstract}
 Supersymmetric quantum Hall liquids are constructed on a noncommutative superplane. 
 We explore a supersymmetric formalism of the Landau problem.
In the lowest Landau level, 
there appear spin-less bosonic states and spin-1/2 
down fermionic states, which exhibit a super-chiral property. 
It is  shown the Laughlin wavefunction and
 topological excitations have their superpartners.
Similarities  between  supersymmetric quantum Hall systems
 and bilayer quantum Hall systems are discussed.  
\end{abstract}

\maketitle

%%%%%%%%%%%%%%%%%%%%%%%%%%%%%%%%%%%%%%%%%%%%%%%%%%%%%%%%%
%%%%%%%%%%%%%%%%%%%%%%%%%%%%%%%%%%%%%%%%%%%%%%%%%%%%%%%%%%
%%%%%%%%%%%%%%%%%%%%%%%%%%%%%%%%%
\section{Introduction}
%%%%%%%%%%%%%%%%%%%%%%%%%%%%%%%%%%%%%%%%%%%%%%%%%%%%%%%%%
%%%%%%%%%%%%%%%%%%%%%%%%%%%%%%%%%%%%%%%%%%%%%%%%%%%%%%%%%%
%%%%%%%%%%%%%%%%%%%%%%%%%%%%%%%%%

 Over the past few years, accompanied with the developments of the noncommutative (NC) geometry  and string theory, quantum Hall (QH)
 systems 
 have attracted increasing attentions from particle physicists. 
(See \cite{hep-th/0101029,hep-th/0103013}
 for instance.) 
 It is well known that 
 the underlying mathematical structure  of  QH system 
 is  NC geometry, and QH systems manifest its exotic properties
 \cite{GirvinPhysRevB29,IsoPLB92,cond-mat/9407031,hep-th/0209198}.
  Based on the second Hopf map,    
 a four dimensional generalization of QH liquid  
was constructed in Ref. \cite{cond-mat/0110572}. 
 The system 
  has higher dimensional analogues of the exotic structures of 
 the two dimensional QH system, such as NC geometry, fractionally 
 charged excitations,  massless edge states, etc. 
Since then, many efforts are devoted to the understanding of the 
 four dimensional  QH liquid 
\cite{HuPRB66,hep-th/0204256,hep-th/0210059,cond-mat/0211679,hep-th/0203095,
hep-th/0209104,cond-mat/0305697,cond-mat/0206164,hep-th/0306277} and the
 construction of 
even higher dimensional QH systems
  \cite{hep-th/0203264,hep-th/0201016,cond-mat/0306045,cond-mat/0306351,hasebekimura0310274}.
 The studies of 
 higher dimensional QH systems have brought 
 many fruitful developments in both particle physics and condensed matter physics.
Particularly, spherical boundstates of 
 D-branes in string theory were well investigated 
  based 
 on the set-up of the fuzzy spheres in
  higher dimensional QH systems \cite{hep-th/0402044} .
 Three dimensional reduction of four dimensional QH effect
   gave a hint to the discovery of 
 the spin-Hall effect \cite{Murakamiscience}, which has become one of 
 the most rapidly growing topics in condensed matter physics.

 Recently, it was discovered that the non-anticommutative (NAC) 
   field theory is naturally realized on D-branes in Ramond-Ramond field or 
 graviphoton background \cite{hep-th/0302109,hep-th/0302078, hep-th/0305248, 
hep-th/0311005}.  
Also, it has been shown that, in the supermatrix model, 
fluctuations on a fuzzy supersphere yield supersymmetric  
 NC field theories \cite{hep-th/0311005}. 
 Besides, some interesting relations between  
  NAC geometry,  Landau problem and 
  QH systems are reported \cite{hep-th/0306251,hep-th/0311159,hep-th/0410070,hasebekimura0409230}.   
Especially, on a fuzzy supersphere, a supersymmetric extension of QH  
 liquid  was explicitly constructed in Ref.\cite{HasebePRL29}.
While   mathematical properties of  NAC theories 
have been well investigated \cite{hep-th/0306215,
hep-th/0307039,hep-th/0307076}, 
 their emergent physical consequences have  not been  satisfactorily 
 understood yet. 
The supersymmetric QH system provides a rare ``physical''
 set-up whose underlying mathematics  is given by NAC geometry.   
  Since two dimensional and higher dimensional 
 QH systems  manifest peculiar properties of  NC geometry,
 it would be reasonable to  expect that  
 explorations of supersymmetric QH liquids
 may reveal yet unknown physical aspects of the NAC
 geometry. 
In this paper, by taking a planar limit of the fuzzy 
{supersphere}, 
we construct QH liquids on a NC 
 superplane, and  investigate 
  physical properties   in a NAC world.

 This paper is organized as follows.
In Section \ref{SuperNCplane}, we review a  systematic construction of   NC superplane
 from the fuzzy supersphere.
It is shown that 
the NC superplane is realized by introducing 
 the super gauge fields. 
In Section \ref{Lagrangeformalism},
we develop Lagrangian and  Hamiltonian formalisms for one-particle system on the 
NC superplane. 
The system possesses (complex) $\mathcal{N}=2$ supersymmetry,
 one of which is   dynamical
and the other is  non-dynamical. 
 Another approach for  one-particle system on a NC plane 
with supersymmetry is found in Ref.\cite{hep-th/0405271,hep-th/0407247}, where 
a higher derivative term is introduced    
 to be invariant under the Galilean boosts transformation. 
% %
In Section \ref{Hilbertspace}, 
we analyze a supersymmetric Landau problem.
In each of the higher Landau levels (LLs), there exists $\mathcal{N}=2$ supersymmetry,
 while in the lowest Landau level (LLL), only the $\mathcal{N}=1$
 non-dynamical supersymmetry 
 remains valid.
We explicitly construct radially symmetric orbit
 states, which form a ``complete'' basis in the LLL. 
These states are   super-holomorphic
 except for their exponential term, and show a super-chiral 
property where not only the orbital rotation but also 
the spin polarization is chiral.
In Section \ref{Laughlinplane}, 
 a Laughlin wavefunction and its superpartner on 
the NC superplane are derived.
In Section \ref{excitedplane},
we present bosonic and fermionic  
topological excitations, and investigate their basic properties.
In Section \ref{realsystem},
 we  discuss a possible mapping from  supersymmetric QH 
 systems to bilayer QH systems.
Section \ref{summary}
 is devoted for summary and discussions.
%
%
%In Appendix \ref{magnetictrans},
%  magnetic translations  on the superplane and their accompanied 
%Aharonov-Bohm phases are summarized.
%
%In Appendix \ref{infinitsym}, a supersymmetric extension of the 
%$W_{\infty}$ algebra in the LLL
% is discussed.
%
%In Appendix \ref{radiallysymorbits}, 
%the explicit expression of the basis for  
%the  supersymmetric Landau problem in the symmetric gauge 
%is presented.
%
%In Appendix \ref{Neumann}, 
% we concisely give the von Neumann formalism on a superplane.

%%%%%%%%%%%%%%%%%%%%%%%%%%%%%%%%%%%%%%%%%%%%%%%%%%%%%%%%%%%%%%%%%
%%%%%%%%%%%%%%%%%%%%%%%%%%%%%%%%%%%%%%%%%%%%%%%%%%%%%%%%%%%%%%%%%
%%%%%%%%%%%%%%%%%%
%%%%%%%%%%%%%%%%%%%%%%%%%%%%%%%%%%%%%%%%%%%%%%%%%%%%%%%%%%%%%%%%%%
%%%%%%%%%%%%%%%%%%%%%%%%%%%%%%%%%%%%%%%%%%%%%%%%%%%%%%%%%%%%%%%%%%
%%%%%%%%%%%%%%%

%%%%%%%%%%%%%%%%%%%%%%%%%%%%%%%%%%%%%%%%%%%%%%%%%%%%%%%%%%%%%%%%%%%%%%%%%
%%%%%%%%%%%%%%%%%%%%%%%%%%%%%%%%%%%%%%%%%%%%%%%%%%%%%%%%%%%%%%%%%%%%
\section{Noncommutative superplane}\label{SuperNCplane}
%%%%%%%%%%%%%%%%%%%%%%%%%%%%%%%%%%%%%%%%%%%%%%%%%%%%%%%%%%%%%%%%%%%%%%%%%
%%%%%%%%%%%%%%%%%%%%%%%%%%%%%%%%%%%%%%%%%%%%%%%%%%%%%%%%%%%%%%%%%%%%

 Based on Ref.\cite{hep-th/0306251},
 we review an algebra on a NC  superplane from the 
 $OSp(1|2)$ algebra.
The $OSp(1|2)$ algebra consists of five generators $L_a (a=1,2,3)$ and 
 $L_{\alpha} (\alpha=1,2)$,
%%%%%%%%%%%%%%%%%%%%%%%%%%%%%%%%%%%%%%%%%%%%%%%%%%%%%%%%%%%%%%%%%%%%%%%%%%
\begin{subequations}
\begin{align}
&[L_a,L_b]=i\epsilon_{abc}L_c,\\
&[L_a,L_{\alpha}]=\frac{1}{2}(\sigma_a)_{\beta\alpha}L_{\beta},\\
&\{L_{\alpha},L_{\beta}\}=\frac{1}{2}(C\sigma_a)_{\alpha\beta}L_a,
\end{align}
\end{subequations}
%%%%%%%%%%%%%%%%%%%%%%%%%%%%%%%%%%%%%%%%%%%%%%%%%%%%%%%%%%%%%%%%%%%%%%%%%%
where $\{\sigma_a\}$ are Pauli matrices and $C$ denotes a charge conjugation 
 matrix $C=i\sigma_2$. 
With a given noncommutative scale $\alpha$, the coordinates 
on the fuzzy supersphere $S_F^{2|2}$ is
 identified with the $OSp(1|2)$ generators by 
 $X_a=\alpha L_{a}$ and $\Theta_{\alpha}=\alpha L_{\alpha}$
\cite{hasebekimura0409230}.

We apply  a symmetric scaling to the $OSp(1|2)$ generators as
%%%%%%%%%%%%%%%%%%%%%%%%%%%%%%%%%%%%%%%%%%%%%%%%%%%%%%%%%%%%%%%%%%%%%%%%%
\begin{subequations}\begin{align}
&(L_i,L_{\alpha})\rightarrow (T_i,T_{\alpha})=\epsilon (L_i,L_{\alpha}),\\
&L_3 \rightarrow L_{\perp},
\end{align}\end{subequations} 
%%%%%%%%%%%%%%%%%%%%%%%%%%%%%%%%%%%%%%%%%%%%%%%%%%%%%%%%%%%%%%%%%%%%%%%%%
where $i=1,2$.
By taking the limit $\epsilon \rightarrow 0$, the $OSp(1|2)$ algebra  
 reduces to the translation and rotation algebras on the superplane
%%%%%%%%%%%%%%%%%%%%%%%%%%%%%%%%%%%%%%%%%%%%%%%%%%%%%%%%%%%%%%%%%%%%%%%%%
\begin{subequations}\begin{align}
&[T_i,T_j]=0,~[T_i,L_{\perp}]=-i\epsilon_{ij}T_j,\label{su2}\\
& [T_i,T_{\alpha}]=0, \label{decoup}\\
&\{T_{\alpha},T_{\beta}\}=0,~[T_{\alpha},L_{\perp}]=\pm 
\frac{1}{2}T_{\alpha}, \label{fsu2}
\end{align}\label{algebrasuperplane}\end{subequations}
%%%%%%%%%%%%%%%%%%%%%%%%%%%%%%%%%%%%%%%%%%%%%%%%%%%%%%%%%%%%%%%%%%%%%%%%%
where,  in Eq.(\ref{fsu2}), $+$ corresponds to $\alpha=\theta_1$, and 
$-$ corresponds to $\alpha=\theta_2$.
Eq.(\ref{su2}) represents the algebra of the two dimensional  
Euclidean group. Similarly, Eq.(\ref{fsu2}) may be regarded as
 the algebra of the symmetry group on the two dimensional fermionic plane. 
The differential  representation  for the algebras 
(\ref{algebrasuperplane}) is  given by
%%%%%%%%%%%%%%%%%%%%%%%%%%%%%%%%%%%%%%%%%%%%%%%%%%%%%%%%%%%%%%%%%%%%%%%%%   
\begin{subequations}\begin{align}
&T_i=-i\partial_i,~~T_{\alpha}=-i\partial_{\alpha},\\
&L_{\perp}=(\sigma_2)_{ij}x_i\partial_j
+\frac{1}{2}(\sigma_3)_{\alpha\beta}\theta_{\alpha}\partial_{\beta}.
\label{angularplane}
\end{align}\end{subequations}
%%%%%%%%%%%%%%%%%%%%%%%%%%%%%%%%%%%%%%%%%%%%%%%%%%%%%%%%%%%%%%%%%%%%%%%%%%

Around the north pole on the fuzzy supersphere,
 $X_3\sim \alpha  j$ 
(where $j$ is a superspin index which specifies  irreducible 
representations of the $OSp(1|2)$ group), 
 the  NC algebras   on the fuzzy supersphere 
 reduce to those  on the NC superplane $R_{NC}^{2|2}$,
%%%%%%%%%%%%%%%%%%%%%%%%%%%%%%%%%%%%%%%%%%%%%%%%%%%%%%%%%%%%%%%%%%%%%%%%%
\begin{subequations}\begin{align}
&[\hat{X}_1,\hat{X}_2]= -i,\label{xbsu2}\\
& [\hat{X}_i,\hat{\Theta}_{\alpha}]= 0, \label{xdecoup}\\
&\{\hat{\Theta}_{1},\hat{\Theta}_{2}\}=1,\label{xfsu2}
\end{align}\label{xalgebrasuperplane}\end{subequations}
%%%%%%%%%%%%%%%%%%%%%%%%%%%%%%%%%%%%%%%%%%%%%%%%%%%%%%%%%%%%%%%%%%%%%%%%%
where we have defined the dimensionless coordinates  
as $\hat{X_i}=\frac{1}{\alpha \sqrt{j}}X_i,~\hat{\Theta}_{\alpha}
=\frac{\sqrt{2}}{\alpha \sqrt{j}}\Theta_{\alpha}$. 
(More general contractions, 
 including asymmetric scaling, are found in Ref.\cite{hep-th/0306251}.)
The bosonic coordinates and the fermionic coordinates  are 
completely decoupled  unlike the fuzzy supersphere 
case. 
The  algebra (\ref{xbsu2})  
is equivalent to that on the NC bosonic plane.
The original QH systems on  NC bosonic plane  
  have  already been  well investigated as found in Ref.{\cite{BookEzawa}}.
 In the following, we include the known results on the bosonic NC plane for complete description.
%%%%%%%%%%%%%%%%%%%%%%%%%%%%%%%%%%%%%%%%%%%%%%%%%%%%%%%%%%%%%%%%%%%%%%%%%
%\subsection{Noncommutative Algebra in Superplanar QH Systems}
%\label{superNCplane}
%%%%%%%%%%%%%%%%%%%%%%%%%%%%%%%%%%%%%%%%%%%%%%%%%%%%%%%%%%%%%%%%%%%%%%%%%%

A physical set-up for the NC superplane is realized by introducing 
 super gauge fields.
We consider a constant magnetic strength made by a 
bosonic gauge field and a fermionic gauge field as
%%%%%%%%%%%%%%%%%%%%%%%%%%%%%%%%%%%%%%%%%%%%%%%%%%%%%%%%%%%%%%%%%%%%%%%%%%
\begin{subequations}\begin{align}
&B=-i(\sigma_2)_{ij}\partial_i A_j=-\epsilon_{ij}\partial_iA_j,
\label{BfromAi}\\
&B=-i(\sigma_3)_{\alpha\beta}\partial_{\alpha}(C_{\beta\gamma}A_{\gamma})
=-i(\sigma_1)_{\alpha\beta}\partial_{\alpha}A_{\beta}.
\label{BfromAalpha}
\end{align}\end{subequations}\label{constmagbf}
%%%%%%%%%%%%%%%%%%%%%%%%%%%%%%%%%%%%%%%%%%%%%%%%%%%%%%%%%%%%%%%%%%%%%%%%%% 
It is apparent that there exists a  $U(1)$ gauge degree of freedom,
%%%%%%%%%%%%%%%%%%%%%%%%%%%%%%%%%%%%%%%%%%%%%%%%%%%%%%%%%%%%%%%%%%%%%%%%%%
$A_i\rightarrow A_i+\partial_i \xi$ and 
 $A_{\alpha}\rightarrow A_{\alpha}+\partial_{\alpha}\xi$.
%%%%%%%%%%%%%%%%%%%%%%%%%%%%%%%%%%%%%%%%%%%%%%%%%%%%%%%%%%%%%%%%%%%%%%%%%% 
The covariant momenta are given by
%%%%%%%%%%%%%%%%%%%%%%%%%%%%%%%%%%%%%%%%%%%%%%%%%%%%%%%%%%%%%%%%%%%%%%%%%%
\begin{subequations}\begin{align}
&P_i=-i(\partial_i+iA_i),\label{covmomi}\\
&P_{\alpha}=i(\partial_{\alpha}+iA_{\alpha}).\label{covmomalpha}
\end{align}\label{covmom}\end{subequations}
%%%%%%%%%%%%%%%%%%%%%%%%%%%%%%%%%%%%%%%%%%%%%%%%%%%%%%%%%%%%%%%%%%%%%%%%%% 
With these covariant momenta, the center-of-mass coordinates are 
defined as
%%%%%%%%%%%%%%%%%%%%%%%%%%%%%%%%%%%%%%%%%%%%%%%%%%%%%%%%%%%%%%%%%%%%%%%%%%
\begin{subequations}\begin{align}
&X_i=x_i+i\ell_B^2(\sigma_2)_{ij}P_j, \label{centerofmassx}\\
&\Theta_{\alpha}=\theta_{\alpha}-i\ell_B^2(\sigma_1)_{\alpha\beta}P_{\beta},
 \label{centerofmasstheta}
\end{align}\label{centerofmass}
\end{subequations} 
%%%%%%%%%%%%%%%%%%%%%%%%%%%%%%%%%%%%%%%%%%%%%%%%%%%%%%%%%%%%%%%%%%%%%%%%%%
where $\ell_B\equiv 1/\sqrt{B}$ is the magnetic length.

The center-of-mass coordinates and the covariant momenta are 
completely decoupled, and  satisfy the super Heisenberg-Weyl algebra 
individually,
%%%%%%%%%%%%%%%%%%%%%%%%%%%%%%%%%%%%%%%%%%%%%%%%%%%%%%%%%%%%%%%%%%%%%%%%%
\begin{subequations}\begin{align}
&[P_i,P_j]=-\frac{1}{\ell_B^2}(\sigma_2)_{ij},\\
&[P_i,P_{\alpha}]=0,\\
&\{P_{\alpha},P_{\beta}\}=\frac{1}{\ell_B^2}(\sigma_1)_{\alpha\beta},
\end{align}\end{subequations}
%%%%%%%%%%%%%%%%%%%%%%%%%%%%%%%%%%%%%%%%%%%%%%%%%%%%%%%%%%%%%%%%%%%%%%%%%%
and 
%%%%%%%%%%%%%%%%%%%%%%%%%%%%%%%%%%%%%%%%%%%%%%%%%%%%%%%%%%%%%%%%%%%%%%%%%
\begin{subequations}\begin{align}
&[X_i,X_j]=\ell_B^2 (\sigma_2)_{ij},\label{supercommx}\\ 
&[X_i,\Theta_{\alpha}]=0,\label{xthetadecoup}\\
&\{\Theta_{\alpha},\Theta_{\beta}\}= 
\ell_B^2(\sigma_1)_{\alpha\beta}.\label{supercommtheta} 
\end{align}\label{totalsupercommx}\end{subequations}
%%%%%%%%%%%%%%%%%%%%%%%%%%%%%%%%%%%%%%%%%%%%%%%%%%%%%%%%%%%%%%%%%%%%%%%%
The set of  algebras (\ref{totalsupercommx}) is 
 consistent with Eq.(\ref{xalgebrasuperplane}).
In the LLL limit ($B\rightarrow \infty$),  it is easily seen from 
Eq.(\ref{centerofmass}) the particle position $(x_i, \theta_{\alpha})$ 
 reduces to the center-of-mass coordinate operator $(X_i,\Theta_{\alpha})$, 
and the superplane under the strong super magnetic field is identified with the 
NC superplane. 

The angular momentum (\ref{angularplane}) can be rewritten in terms of 
the covariant momenta and the center-of-mass coordinates  as
%%%%%%%%%%%%%%%%%%%%%%%%%%%%%%%%%%%%%%%%%%%%%%%%%%%%%%%%%%%%%%%%%%%%%%%%%
\begin{equation}
L_{\perp}=\frac{1}{2\ell_B^2}(X_i^2+\frac{1}{2}C_{\alpha\beta}
\Theta_{\alpha}\Theta_{\beta})-\frac{1}{2}\ell_B^2(P_i^2+
\frac{1}{2}C_{\alpha\beta}P_{\alpha}
P_{\beta}).
%L_{\perp}^B+L_{\perp}^F,
\end{equation}
%%%%%%%%%%%%%%%%%%%%%%%%%%%%%%%%%%%%%%%%%%%%%%%%%%%%%%%%%%%%%%%%%%%%%%%%%%
%%%%%%%%%%%%%%%%%%%%%%%%%%%%%%%%%%%%%%%%%%%%%%%%%%%%%%%%%%%%%%%%%%%%%%%%%%
The center-of-mass coordinates
 $(X_i,\Theta_{\alpha})$ and  the covariant momenta $(P_i,P_{\alpha})$ 
form a closed algebra with $L_{\perp}$, individually, 
%%%%%%%%%%%%%%%%%%%%%%%%%%%%%%%%%%%%%%%%%%%%%%%%%%%%%%%%%%%%%%%%%%%%%%%%%
\begin{subequations}\begin{align}
&[L_{\perp},X_i]=-(\sigma_2)_{ij}X_j,~~[L_{\perp},\Theta_{\alpha}]
=\frac{1}{2}(\sigma_3)_{\alpha\beta}\Theta_{\beta},\\
&[L_{\perp},P_i]=(\sigma_2)_{ij}P_j,~~~~[L_{\perp},P_{\alpha}]
=-\frac{1}{2}(\sigma_3)_{\alpha\beta}P_{\beta}.
\end{align}\end{subequations}
%%%%%%%%%%%%%%%%%%%%%%%%%%%%%%%%%%%%%%%%%%%%%%%%%%%%%%%%%%%%%%%%%%%%%%%%%

Due to  the existence of  two sets of the super Heisenberg-Weyl algebras, 
  two sets of  supersymmetric harmonic oscillators are naturally defined.
The bosonic creation   and   annihilation operators 
  are given by   
%%%%%%%%%%%%%%%%%%%%%%%%%%%%%%%%%%%%%%%%%%%%%%%%%%%%%%%%%%%%%%%%%%%%%%%%%%
\begin{subequations}\begin{align}
&a\equiv \frac{\ell_B}{\sqrt{2}}(P_x+iP_y), 
~~a^{\dagger}\equiv\frac{\ell_B}{\sqrt{2}}(P_x-iP_y), \\
&b\equiv\frac{1}{\sqrt{2}\ell_B}(X-iY),
 ~~b^{\dagger}\equiv\frac{1}{\sqrt{2}\ell_B}(X+iY), \label{bbdagger}
\end{align}\label{bosoniccreanni}
\end{subequations}
%%%%%%%%%%%%%%%%%%%%%%%%%%%%%%%%%%%%%%%%%%%%%%%%%%%%%%%%%%%%%%%%%%%%%%%%%
which satisfy 
%%%%%%%%%%%%%%%%%%%%%%%%%%%%%%%%%%%%%%%%%%%%%%%%%%%%%%%%%%%%%%%%%%%%%%%%%
$[a,a^{\dagger}]=[b,b^{\dagger}]=1$.
Other commutators become  zeros.
Similarly, the fermionic creation  and  annihilation operators are given by 
%%%%%%%%%%%%%%%%%%%%%%%%%%%%%%%%%%%%%%%%%%%%%%%%%%%%%%%%%%%%%%%%%%%%%%%%%%
\begin{subequations}\begin{align}
&\alpha\equiv\ell_B P_{\theta_2},~~\alpha^{\dagger}\equiv\ell_B P_{\theta_1},\\
&\beta\equiv\frac{1}{\ell_B} \Theta_{2},~~\beta^{\dagger}\equiv\frac{1}{\ell_B}
\Theta_1,\label{betabetadagger}
\end{align}\label{fermicreanni}\end{subequations}
%%%%%%%%%%%%%%%%%%%%%%%%%%%%%%%%%%%%%%%%%%%%%%%%%%%%%%%%%%%%%%%%%%%%%%%%%%
which satisfy 
$\{\alpha,\alpha^{\dagger}\}=\{\beta,\beta^{\dagger}\}=1$.
Other anticommutators  are zeros.
With use of  supersymmetric harmonic oscillators, 
the angular momentum can be written as 
%%%%%%%%%%%%%%%%%%%%%%%%%%%%%%%%%%%%%%%%%%%%%%%%%%%%%%%%%%%%%%%%%%%%%%%%%
\begin{equation}
L_{\perp}=(b^{\dagger}b+\frac{1}{2}\beta^{\dagger}
\beta)-(a^{\dagger}a+\frac{1}{2}\alpha^{\dagger}\alpha).
\end{equation}
%%%%%%%%%%%%%%%%%%%%%%%%%%%%%%%%%%%%%%%%%%%%%%%%%%%%%%%%%%%%%%%%%%%%%%%%%%
Thus, the  $b$-quantum 
acquires the angular momentum by $1$, while the  $\beta$-quantum 
acquires the angular momentum by $1/2$.

%%%%%%%%%%%%%%%%%%%%%%%%%%%%%%%%%%%%%%%%%%%%%%%%%%%%%%%%%%%%%%%%%%%%%
%%%%%%%%%%%%%%%%%%%%%%%%%%%%%%%%%%%%%%%%%%%%%%%%%%%%%%%%%%%%%%%%%%%%%%%%
%\subsection{Symmetric Gauge}
%%%%%%%%%%%%%%%%%%%%%%%%%%%%%%%%%%%%%%%%%%%%%%%%%%%%%%%%%%%%%%%%%%%%%%%
%%%%%%%%%%%%%%%%%%%%%%%%%%%%%%%%%%%%%%%%%%%%%%%%%%%%%%%%%%%%%%%%%%%%%%

It is convenient to  fix the gauge freedom as the symmetric gauge,
%%%%%%%%%%%%%%%%%%%%%%%%%%%%%%%%%%%%%%%%%%%%%%%%%%%%%%%%%%%%%%%%%%%%%%%%%
\begin{equation}
A_i=i(\sigma_2)_{ij}x_j\frac{B}{2},
~~A_{\alpha}=i(\sigma_1)_{\alpha\beta}\theta_{\beta}\frac{B}{2}.
\end{equation}
%%%%%%%%%%%%%%%%%%%%%%%%%%%%%%%%%%%%%%%%%%%%%%%%%%%%%%%%%%%%%%%%%%%%%%%%%
These expressions  
 are obtained by  expanding the 
 supermonopole gauge fields \cite{hasebekimura0409230}
 around the north pole on the supersphere.
The field strengths become
%%%%%%%%%%%%%%%%%%%%%%%%%%%%%%%%%%%%%%%%%%%%%%%%%%%%%%%%%%%%%%%%%%%%%%%%%%
\begin{subequations}\begin{align}
&F_{ij}=\partial_{i}A_j-\partial_j A_i=-iB(\sigma_2)_{ij},
\label{planebosonicfield}\\
&F_{i\alpha}=\partial_i A_{\alpha}-\partial_{\alpha}A_i=0,
\label{planebosonfermionfield}\\
&F_{\alpha\beta}=\partial_{\alpha}A_{\beta}+\partial_{\beta}A_{\alpha}=
iB(\sigma_1)_{\alpha\beta}.\label{planefermionicfield}
\end{align}\end{subequations}
%%%%%%%%%%%%%%%%%%%%%%%%%%%%%%%%%%%%%%%%%%%%%%%%%%%%%%%%%%%%%%%%%%%%%%%%%%
In the symmetric gauge, the  creation and annihilation 
operators (\ref{bosoniccreanni}) (\ref{fermicreanni}) read  as
%%%%%%%%%%%%%%%%%%%%%%%%%%%%%%%%%%%%%%%%%%%%%%%%%%%%%%%%%%%%%%%%%%%%%%%%%
\begin{subequations}\begin{align}
&a=-\frac{i}{\sqrt{2}}(z+\partial^*),~~a^{\dagger}=
\frac{i}{\sqrt{2}}(z^*-\partial),\\
&b=\frac{1}{\sqrt{2}}(z^*+\partial),~~b^{\dagger}=
\frac{1}{\sqrt{2}}(z-\partial^*),
\end{align}\end{subequations}
%%%%%%%%%%%%%%%%%%%%%%%%%%%%%%%%%%%%%%%%%%%%%%%%%%%%%%%%%%%%%%%%%%%%%%%%%
and 
%%%%%%%%%%%%%%%%%%%%%%%%%%%%%%%%%%%%%%%%%%%%%%%%%%%%%%%%%%%%%%%%%%%%%%%%%
\begin{subequations}\begin{align}
&\alpha=-\frac{i}{\sqrt{2}}(\theta-\partial_{\theta}^*),~~\alpha^{\dagger}
=-\frac{i}{\sqrt{2}}(\theta^*-\partial_{\theta}),\\
&\beta=\frac{1}{\sqrt{2}}(\theta^*+\partial_{\theta}),~~\beta^{\dagger}
=\frac{1}{\sqrt{2}}(\theta+\partial_{\theta}^*),
\end{align}\end{subequations}
%%%%%%%%%%%%%%%%%%%%%%%%%%%%%%%%%%%%%%%%%%%%%%%%%%%%%%%%%%%%%%%%%%%%%%%%%%
where we have used  dimensionless complex coordinates and 
  derivatives,   
%%%%%%%%%%%%%%%%%%%%%%%%%%%%%%%%%%%%%%%%%%%%%%%%%%%%%%%%%%%%%%%%%%%%%%%%%
\begin{subequations}\begin{align}
&z=\frac{1}{2\ell_B}(x+iy),~~z^{*}=\frac{1}{2\ell_B}(x-iy),\\
&\partial=\ell_B(\partial_x-i\partial_y),~~\partial^*=
\ell_B(\partial_x+i\partial_y),
\end{align}\end{subequations}
%%%%%%%%%%%%%%%%%%%%%%%%%%%%%%%%%%%%%%%%%%%%%%%%%%%%%%%%%%%%%%%%%%%%%%%%
and 
%%%%%%%%%%%%%%%%%%%%%%%%%%%%%%%%%%%%%%%%%%%%%%%%%%%%%%%%%%%%%%%%%%%%%%%%%
\begin{subequations}\begin{align}
&\theta=\frac{1}{\sqrt{2}\ell_B}\theta_1,~~\theta^{*}
=\frac{1}{\sqrt{2}\ell_B}\theta_2,\\
&\partial_{\theta}=\sqrt{2}\ell_B\partial_{\theta_1},
~~\partial_{\theta}^*=\sqrt{2}\ell_B\partial_{\theta_2}.
\end{align}\end{subequations}
%%%%%%%%%%%%%%%%%%%%%%%%%%%%%%%%%%%%%%%%%%%%%%%%%%%%%%%%%%%%%%%%%%%%%%%%%

%%%%%%%%%%%%%%%%%%%%%%%%%%%%%%%%%%%%%%%%%%%%%%%%%%%%%%%%%%%%%%%%%%%%%%%%%%
%%%%%%%%%%%%%%%%%%%%%%%%%%%%%%%%%%%%%%%%%%%%%%%%%%%%%%%%%%%%%%%%%%%
\section{ one-particle Hamiltonian and supersymmetry }\label{Lagrangeformalism}
%%%%%%%%%%%%%%%%%%%%%%%%%%%%%%%%%%%%%%%%%%%%%%%%%%%%%%%%%%%%%%%%%%%%%%%%%
%%%%%%%%%%%%%%%%%%%%%%%%%%%%%%%%%%%%%%%%%%%%%%%%%%%%%%%%%%%%%%%%%%%%

We develop a Lagrangian formalism for  one-particle 
 in the presence of super gauge fields.
The Lagrangian  may be given by 
%\subsection{Lagrangean}
%%%%%%%%%%%%%%%%%%%%%%%%%%%%%%%%%%%%%%%%%%%%%%%%%%%%%%%%%%%%%%%%%%%%%%%%%%
\begin{equation}
L=\frac{M}{2}(\dot{x}_i^2+C_{\alpha\beta}\dot{\theta}_{\alpha}
\dot{\theta}_{\beta})-(A_i\dot{x}_i+A_{\alpha}\dot{\theta}_{\alpha}).
\label{lagrangeforplanar}
\end{equation}
%%%%%%%%%%%%%%%%%%%%%%%%%%%%%%%%%%%%%%%%%%%%%%%%%%%%%%%%%%%%%%%%%%%%%%%%%%
In the LLL limit, the kinetic term is quenched, and the 
 Lagrangian (\ref{lagrangeforplanar})  reduces to  
%%%%%%%%%%%%%%%%%%%%%%%%%%%%%%%%%%%%%%%%%%%%%%%%%%%%%%%%%%%%%%%%%%%%%%%%%
\begin{equation}
L_{eff}=-   A_i\dot{x}_i-A_{\alpha}\dot{\theta}_{\alpha}.
\end{equation}
%%%%%%%%%%%%%%%%%%%%%%%%%%%%%%%%%%%%%%%%%%%%%%%%%%%%%%%%%%%%%%%%%%%%%%%%%%
The canonical momenta are derived as   
%%%%%%%%%%%%%%%%%%%%%%%%%%%%%%%%%%%%%%%%%%%%%%%%%%%%%%%%%%%%%%%%%%%%%%%%%%
\begin{subequations}\begin{align}
&p_i= \frac{\partial}{\partial \dot{x}_i}L_{eff}=-A_i=-i(\sigma_2)_{ij}
x_j\frac{B}{2},\\
&p_{\alpha}=\frac{\partial}{\partial \dot{\theta}_{\alpha}}L_{eff}
=A_{\alpha}=i(\sigma_1)_{\alpha\beta}\theta_{\beta}\frac{B}{2},
\end{align}\end{subequations}
%%%%%%%%%%%%%%%%%%%%%%%%%%%%%%%%%%%%%%%%%%%%%%%%%%%%%%%%%%%%%%%%%%%%%%%%%%
where the symmetric gauge was used in the last equations.
By imposing the  commutation relations to canonical variables
%%%%%%%%%%%%%%%%%%%%%%%%%%%%%%%%%%%%%%%%%%%%%%%%%%%%%%%%%%%%%%%%%%%%%%%%%%
\begin{subequations}\begin{align}
&[x_i,p_j]=i\delta_{ij},\\
&\{\theta_{\alpha},p_{\beta}\}=i\delta_{\alpha\beta},
\end{align}\label{useualcanonicalcommu}\end{subequations}
%%%%%%%%%%%%%%%%%%%%%%%%%%%%%%%%%%%%%%%%%%%%%%%%%%%%%%%%%%%%%%%%%%%%%%%%%%
we obtain the NC relations  
%%%%%%%%%%%%%%%%%%%%%%%%%%%%%%%%%%%%%%%%%%%%%%%%%%%%%%%%%%%%%%%%%%%%%%%%%%
\begin{subequations}\begin{align}
&[x_i,x_j]=\ell_B^2(\sigma_2)_{ij},\\
&\{\theta_{\alpha},\theta_{\beta}\}=\ell_B^2(\sigma_1)_{\alpha\beta}.
\end{align}\end{subequations}
%%%%%%%%%%%%%%%%%%%%%%%%%%%%%%%%%%%%%%%%%%%%%%%%%%%%%%%%%%%%%%%%%%%%%%%%%%
These relations are what we have already obtained in 
Eq.(\ref{totalsupercommx}).
Then, it would be reasonable to adopt Eq.(\ref{lagrangeforplanar}) as the 
Lagrangian for the present system.

The equations of motions are derived as 
%%%%%%%%%%%%%%%%%%%%%%%%%%%%%%%%%%%%%%%%%%%%%%%%%%%%%%%%%%%%%%%%%%%%%%%%%
\begin{subequations}\begin{align}
&M\ddot{x}_i=\epsilon_{ij}B\dot{x}_j,\label{eomx}\\
&M\ddot{\theta}_{\alpha}=-i(\sigma_3)_{\alpha\beta}B\dot{\theta}_{\beta},
\label{eomtheta}
\end{align}\label{totaleom}\end{subequations}
%%%%%%%%%%%%%%%%%%%%%%%%%%%%%%%%%%%%%%%%%%%%%%%%%%%%%%%%%%%%%%%%%%%%%%%%%
which represent   cyclotron motions for bosonic  and  
fermionic degrees of freedom. 
As we shall discuss in the next section, the fermionic 
 variables $\{\theta_{\alpha}\}$ are  related to the 
spin degrees of freedom. 
With the definition of the spin 
$S_a=-i\frac{M}{2}\theta_{\alpha}(\sigma_aC)_{\alpha\beta}
\dot{\theta}_{\beta}$, Eq.(\ref{eomtheta}) implies the spin 
precession motion, 
%%%%%%%%%%%%%%%%%%%%%%%%%%%%%%%%%%%%%%%%%%%%%%%%%%%%%%%%%%%%%%%
\begin{equation}
\dot{S}_i=-\epsilon_{ij}S_j B.
\end{equation}
%%%%%%%%%%%%%%%%%%%%%%%%%%%%%%%%%%%%%%%%%%%%%%%%%%%%%%%%%%%%%%%%%
The Lagrangian (\ref{lagrangeforplanar}) apparently possesses    
translational symmetries  on both the bosonic plane and the fermionic 
plane. 
The   Noether charges accompanied by the translational symmetries are 
 obtained as   
%%%%%%%%%%%%%%%%%%%%%%%%%%%%%%%%%%%%%%%%%%%%%%%%%%%%%%%%%%%%%%%%%%%%%%%%%
\begin{subequations}\begin{align}
&\mathcal{P}_i=M\dot{x}-B\epsilon_{ij}x_j,\\
&\mathcal{P}_{\alpha}=MC_{\alpha\beta}\dot{\theta}_{\beta}+
iB(\sigma_1)_{\alpha\beta}\theta_{\beta},
\end{align}\label{planartotalmomenta}\end{subequations}
%%%%%%%%%%%%%%%%%%%%%%%%%%%%%%%%%%%%%%%%%%%%%%%%%%%%%%%%%%%%%%%%%%%%%%%%%%
which are total momenta.
The first terms on the right-hand sides in Eq.(\ref{planartotalmomenta}) 
represent the particle momenta, and the second terms represent the  field momenta.
The total momenta are related to  the center-of-mass coordinates as 
%%%%%%%%%%%%%%%%%%%%%%%%%%%%%%%%%%%%%%%%%%%%%%%%%%%%%%%%%%%%%%%%%%%%%%%%%%
\begin{equation}
\mathcal{P}_i= -B\epsilon_{ij}X_j,~~   \mathcal{P}_{\alpha}=
B (\sigma_1)_{\alpha\beta}\Theta_{\beta}.
\end{equation}
%%%%%%%%%%%%%%%%%%%%%%%%%%%%%%%%%%%%%%%%%%%%%%%%%%%%%%%%%%%%%%%%%%%%%%%%%
Hence, the center-of-mass coordinates are conserved quantities and 
essentially act as
  translational generators on the NC superplane.

Next, we develop a Hamiltonian formalism.
The canonical momenta are given by 
%%%%%%%%%%%%%%%%%%%%%%%%%%%%%%%%%%%%%%%%%%%%%%%%%%%%%%%%%%%%%%%%%%%%%%%%%%
\begin{subequations}\begin{align}
&p_i= \frac{\partial}{\partial \dot{x}_i}L=M\dot{x}_i-A_i,\\
&p_{\alpha}=\frac{\partial}{\partial \dot{\theta}_{\alpha}}L
=MC_{\alpha\beta}\dot{\theta}_{\beta}+A_{\alpha},
\end{align}\end{subequations}
%%%%%%%%%%%%%%%%%%%%%%%%%%%%%%%%%%%%%%%%%%%%%%%%%%%%%%%%%%%%%%%%%%%%%%%%%%
and  Hamiltonian is constructed as 
%%%%%%%%%%%%%%%%%%%%%%%%%%%%%%%%%%%%%%%%%%%%%%%%%%%%%%%%%%%%%%%%%%%%%%%%%%
\begin{equation}
H=\dot{x}_ip_i+\dot{\theta}_{\alpha}p_{\alpha}-L
=\frac{1}{2M}(P_i^2+C_{\alpha\beta}P_{\alpha}P_{\beta}),
\label{hamilsecondaalpha}
\end{equation}
%%%%%%%%%%%%%%%%%%%%%%%%%%%%%%%%%%%%%%%%%%%%%%%%%%%%%%%%%%%%%%%%%%%%%%%%%%
where  
we have used the covariant momenta (\ref{covmom}).

 With use of creation and annihilation operators, 
  two sets of supercharges are naturally defined   as 
%%%%%%%%%%%%%%%%%%%%%%%%%%%%%%%%%%%%%%%%%%%%%%%%%%%%%%%%%%%%%%%%%%%%%%%%%
\begin{subequations}\begin{align}
&Q \equiv a^{\dagger}\alpha,~~Q^{\dagger}  \equiv\alpha^{\dagger}a,\label{qfirst}\\
& \tilde{Q}\equiv  b^{\dagger}\beta,~~ \tilde{Q}^{\dagger}  \equiv\beta^{\dagger}b,
\label{qsecond}
\end{align}\end{subequations}
%%%%%%%%%%%%%%%%%%%%%%%%%%%%%%%%%%%%%%%%%%%%%%%%%%%%%%%%%%%%%%%%%%%%%%%%%%
and the Hamiltonian (\ref{hamilsecondaalpha}) is expressed as
%%%%%%%%%%%%%%%%%%%%%%%%%%%%%%%%%%%%%%%%%%%%%%%%%%%%%%%%%%%%%%%%%%%%%%%%%
\begin{equation}
H=\omega(a^{\dagger}a+\alpha^{\dagger}\alpha)=\omega\{Q,Q^{\dagger}\}.
\label{hamilsecond}
\end{equation}
%%%%%%%%%%%%%%%%%%%%%%%%%%%%%%%%%%%%%%%%%%%%%%%%%%%%%%%%%%%%%%%%%%%%%%%%%
 Thus, the supercharges $(Q,Q^{\dagger})$ generate
a  dynamical supersymmetry. 
This Hamiltonian commutes
 with  four  supercharges, and  the system possesses 
(complex) $\mathcal{N}=2$  supersymmetry.  
Some comments are added here.
The Hamiltonian (\ref{hamilsecond}) is identical to
  the one used in the  one-dimensional  
 supersymmetric  harmonic oscillator system \cite{Nicolai1976}.
However,  the one-dimensional harmonic oscillator 
 system possesses $\mathcal{N}=1$ supersymmetry only,
 while the present system possesses 
$\mathcal{N}=2$ supersymmetry. (See also Sect.\ref{realsystem}.) 
The anticommutator of 
$(\tilde{Q},\tilde{Q}^{\dagger})$ gives 
 the radius  on the NC superplane as
%%%%%%%%%%%%%%%%%%%%%%%%%%%%%%%%%%%%%%%%%%%%%%%%%%%%%%%%%%%%%%%%%%%%%%%%%%
\begin{align} 
&2\ell_B^2\{\tilde{Q},\tilde{Q}^{\dagger}\}=2\ell_B^2(b^{\dagger}b+\beta^{\dagger}\beta)
 \nonumber\\
&~~~~~~~~~~~~~~~~
=X_i^2+C_{\alpha\beta}\Theta_{\alpha}\Theta_{\beta}\equiv R^2 .
\label{radiuscenter}
\end{align}
%%%%%%%%%%%%%%%%%%%%%%%%%%%%%%%%%%%%%%%%%%%%%%%%%%%%%%%%%%%%%%%%%%%%%%%%%%
This  expression implies that the eigenvalue of the radius operator
 $R^2$ takes a semi-positive  value,  
and the supersymmetry generated by  $(\tilde{Q},\tilde{Q}^{\dagger})$ 
is  a $\it{non\!\!-\!\! dynamical}$ one.
Since $R^2$ commutes with the four supercharges,
 $\mathcal{N}=2$ supermultiplet has not only 
 an identical energy but also   
 an identical eigenvalue of the radius operator.
  
The Hamiltonian and the radius operator commute with the angular momentum. 
Then, the four components of the $\mathcal{N}=2$ supermultiplet can be 
taken as  simultaneous eigenstates of the angular momentum.
The angular momentum and the 
supercharges satisfy the commutation relations  
%%%%%%%%%%%%%%%%%%%%%%%%%%%%%%%%%%%%%%%%%%%%%%%%%%%%%%%%%%%%%%%%%%%%%%%%%%
\begin{subequations}\begin{align}
&[L_{\perp},Q]=-\frac{1}{2}Q,~~[L_{\perp},Q^{\dagger}]=\frac{1}{2}
Q^{\dagger},\\
&[L_{\perp},\tilde{Q}]=\frac{1}{2}\tilde{Q},~~[L_{\perp},
\tilde{Q}^{\dagger}]=-\frac{1}{2}\tilde{Q}^{\dagger}.
\end{align}\label{susyangular}
\end{subequations}
%%%%%%%%%%%%%%%%%%%%%%%%%%%%%%%%%%%%%%%%%%%%%%%%%%%%%%%%%%%%%%%%%%%%%%%%%%
Thus, the supersymmetric transformations  change the eigenvalue of the 
angular momentum by $1/2$.

%%%%%%%%%%%%%%%%%%%%%%%%%%%%%%%%%%%%%%%%%%%%%%%%%%%%%%%%%%%%%%%%%%%%%%%%%%%
%%%%%%%%
%%%%%%%%%%%%%%%%%%%%%%%%%%%%%%%%%%%%%%%%%%%%%%%%%%%%%%%%%%%%%%%%%
\section{ Supersymmetric Landau problem }\label{Hilbertspace}
%%%%%%%%%%%%%%%%%%%%%%%%%%%%%%%%%%%%%%%%%%%%%%%%%%%%%%%%%%%%%%%%%%%%%%%%%%

\begin{figure}[tbph]
\includegraphics*[width=90mm]{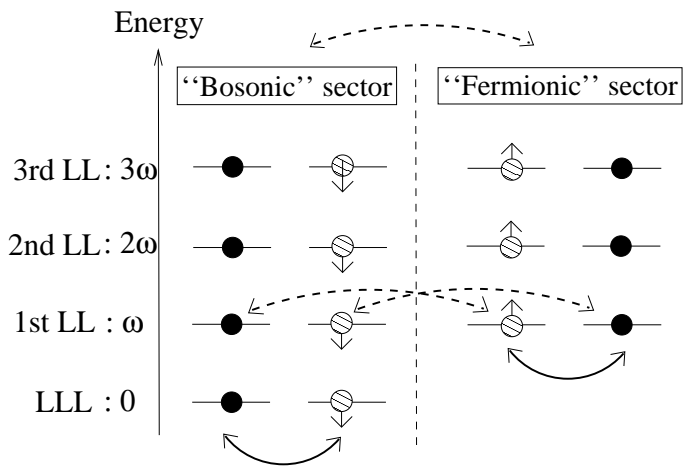}
\caption{  
The left sector about the vertical dashed axis is a ``bosonic sector'' 
 for the dynamical supersymmetry, and the right sector 
 is a ``fermionic sector''.
The curved solid arrows represent the non-dynamical 
supersymmetric transformation generated by $(\tilde{Q},\tilde{Q}^{\dagger})$,
 while the curved dashed arrows represent the dynamical supersymmetry transformation 
generated by $(Q,Q^{\dagger})$. In each of the 
higher LLs,  there are spin-less, spin-1/2 up  and 
spin-1/2 down particles due to the existence of the $\mathcal{N}=2$ 
supersymmetry,
 while, in the LLL, the system possesses only $\mathcal{N}=1$ 
non-dynamical supersymmetry,
 and there appear only spin-less and spin-1/2 down particles. 
}
\label{N2}
\vspace{-3mm}
\end{figure}

The energy spectrum of the Hamiltonian (\ref{hamilsecond}) reads as   
%%%%%%%%%%%%%%%%%%%%%%%%%%%%%%%%%%%%%%%%%%%%%%%%%%%%%%%%%%%%%%%%%%%%%%%%
\begin{equation}
E_n=\omega n,\label{eigensuperenergy}
\end{equation}
%%%%%%%%%%%%%%%%%%%%%%%%%%%%%%%%%%%%%%%%%%%%%%%%%%%%%%%%%%%%%%%%%%%%%%%%%
where $n= 0,1,2,\cdots$ indicates  the LL 
in the supersymmetric 
Landau problem.
[See Appendix \ref{radiallysymorbits} for detail analysis of the eigenvalue 
 problem of the 
 Hamiltonian (\ref{hamilsecond}) and 
 the explicit expression for the eigenstates in the symmetric gauge.]
The zero-point energy is canceled due to the existence of the 
 supersymmetry.
 The higher LLs are doubly degenerate compared to the LLL. 
The eigenvalue of the radius operator  
(\ref{radiuscenter}) is given by  
%%%%%%%%%%%%%%%%%%%%%%%%%%%%%%%%%%%%%%%%%%%%%%%%%%%%%%%%%%%%%%%%%%%%%%%%%%
\begin{equation}
R_m=\sqrt{2m}\ell_B,\label{eigenradiuscenter}
\end{equation}
%%%%%%%%%%%%%%%%%%%%%%%%%%%%%%%%%%%%%%%%%%%%%%%%%%%%%%%%%%%%%%%%%%%%%%%%%%
where $m=0,1,2,\cdots$ indicates the radially symmetric orbits. 
The four components  for  the  $\mathcal{N}=2$ supermultiplet with 
  energy (\ref{eigensuperenergy})  and   radius
 (\ref{eigenradiuscenter}) are constructed as   
%%%%%%%%%%%%%%%%%%%%%%%%%%%%%%%%%%%%%%%%%%%%%%%%%%%%%%%%%%%%%%%%%%%%%%%%%
\begin{subequations}
\begin{align}
& \frac{1}{\sqrt{n!m!}}(a^{\dagger})^n(b^{\dagger})^m|0>, \label{multi1} \\
& \frac{1}{\sqrt{n!(m-1)!}}(a^{\dagger})^{n}\beta^{\dagger}
(b^{\dagger})^{m-1}|0>,\label{multi2} \\
&\frac{1}{\sqrt{(n-1)!m!}}\alpha^{\dagger}(a^{\dagger})^{n-1}
(b^{\dagger})^{m}|0>,\label{multi3} \\
&\frac{1}{\sqrt{(n-1)!(m-1)!}}\alpha^{\dagger}(a^{\dagger})^{n-1}
\beta^{\dagger}(b^{\dagger})^{m-1}|0>.\label{multi4} 
\end{align}\label{supermulti}
\end{subequations} 
%%%%%%%%%%%%%%%%%%%%%%%%%%%%%%%%%%%%%%%%%%%%%%%%%%%%%%%%%%%%%%%%%%%%%%%%
At the same time, they are eigenstates of the  angular momenta $L_{\perp}$ 
with different eigenvalues,
 $l=m-n$, $m-n-\frac{1}{2}$, $m-n+\frac{1}{2}$ 
and $m-n$, respectively. 
Here, we give a physical interpretation of these states. 
Because they have the identical energy  and the radius,
 they may represent four  particle states, 
 which are on the same radially symmetric orbit, and 
 rotate around the origin  with the same frequency.
Hence, they should carry the same $\it{orbital}$ angular momentum,
 while their eigenvalues of the angular momentum $L_{\perp}$ are different.
This discrepancy is solved by noticing that
 $L_{\perp}$ represents the $\it{total}$
 angular momentum, and each of the four particle states
 carries the $\it{intrinsic~ spin}$ as well as the orbital angular momentum.
 Namely, the components of the $\mathcal{N}=2$ supermultiplet 
 (\ref{supermulti})
 are interpreted as the four particle states 
 which have the identical orbital angular momentum $m-n$, and,
  simultaneously, have  different spins
 $0$, $-1/2$, $1/2$ and  $0$, respectively.
 Thus, two of them (\ref{multi1}),(\ref{multi4}) are interpreted as 
 spin-less bosons, and the other 
 two (\ref{multi2}),(\ref{multi3}) are interpreted as 
  spin-1/2 down and up fermions.   
 As suggested by Eq.(\ref{susyangular}), 
 the $\mathcal{N}=2$ supersymmetry changes  their spins by 
 $1/2$, and transforms 
  the bosons to 
   the fermions and vice versa [Fig.\ref{N2}].
%%%%%%%%%%%%%%%%%%%%%%%%%%%%%%%%%%%%%%%%%%%%%%%%%%%%%%%%%%%%%%%%%%
%%%%%%%%%%%%%%%%%%%%%%%%%%%%%%%%%%%%%%%%%%%%%%%%%%%%%%%%%%%%%%%%%5
%%%%%%%%%%%%%%%%%%%%%%%%%%%%%%%%%%%%%%%%%%%%%%%%%%%%%%%%%%%%%%%%%%
It is noted that, in general,
  supersymmetric quantum mechanical models 
 do not deal with a real boson-fermion symmetry \cite{BookJunker},
  while supersymmetric quantum Hall systems
 deal with a real 
 boson-fermion symmetry.  

\begin{figure}[tbph]
\includegraphics*[width=65mm]{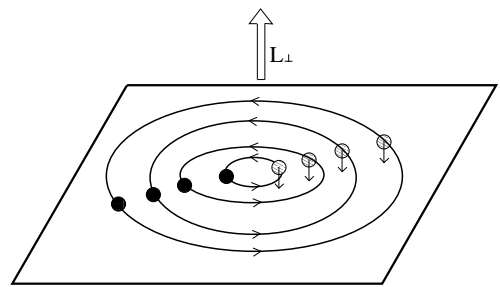}
\caption{ There are  spin-less bosons and  
 spin-1/2 down fermions in the LLL. 
They are on  the radially symmetric orbits,
 and rotate around the origin with the same frequency.  }
\label{correspondingreal.fig}
\vspace{-3mm}
\end{figure}

Each Hilbert space of the higher LL
 possesses the $\mathcal{N}=2$ supersymmetry, 
 because   $n$-th ($n\ge 1$) LL is spanned 
by $\mathcal{N}=2$  supermultiplets (\ref{supermulti})
 with fixed $n$, while, in the LLL, 
 only the non-dynamical supersymmetry $\mathcal{N}=1$ remains valid,
 because the LLL is the ``vacuum'' for 
 the $\mathcal{N}=1$ dynamical supersymmetry. 
In fact,  in the LLL, the Hilbert space 
 is spanned only by the $\mathcal{N}=1$ non-dynamical superpartners 
%%%%%%%%%%%%%%%%%%%%%%%%%%%%%%%%%%%%%%%%%%%%%%%%%%%%%%%%%%%%%%%%%%%%%%%%%%
\begin{subequations}
\begin{align}
&|m+{1}/{2}> =\frac{1}{\sqrt{m!}}\beta^{\dagger}(b^{\dagger})^m|0>, \\
&|m+1>=\frac{1}{\sqrt{(m+1)!}}(b^{\dagger})^{m+1}|0>,
\end{align}\label{otherbasisLLL}
\end{subequations}
%%%%%%%%%%%%%%%%%%%%%%%%%%%%%%%%%%%%%%%%%%%%%%%%%%%%%%%%%%%%%%%%%%%%%
(and the vacuum $|0>$). In the symmetric gauge, with expression of the vacuum 
$\psi_0=\sqrt{\frac{1}{\pi}}e^{-|z|^2-\theta\theta^*}$, 
 they are   represented as 
%%%%%%%%%%%%%%%%%%%%%%%%%%%%%%%%%%%%%%%%%%%%%%%%%%%%%%%%%%%%%%%%%%%%%%%%%
\begin{subequations}
\begin{align}
&\psi_{m+1/2}= \sqrt{\frac{2^{m+1}}{\pi m!}} z^m\theta 
e^{-|z|^2-\theta\theta^*},\\
&\psi_{m+1}  =
\sqrt{\frac{2^{m+1}}{\pi (m+1)!}} z^{m+1} e^{-|z|^2-\theta\theta^*}.
\end{align}
\end{subequations}
%%%%%%%%%%%%%%%%%%%%%%%%%%%%%%%%%%%%%%%%%%%%%%%%%%%%%%%%%%%%%%%%%%%%%%%%
The ``complete  relation'' in the  LLL is obtained as
%%%%%%%%%%%%%%%%%%%%%%%%%%%%%%%%%%%%%%%%%%%%%%%%%%%%%%%%%%%%%%%%%%%%%%%%%%
\begin{align}
&\sum_{m\in 0, \mathbb{N}/2}    \psi_{m}(z,z^*,\theta,\theta^*)
\psi_m^*(z',z'^*,\theta',\theta'^*)   \nonumber\\
&~~~~~~~~~=\frac{1}{\pi}e^{ - (|z|^2+\theta\theta^*) -  
(|z'|^2+\theta'\theta'^*)-2(z'^*z+\theta'^*\theta) }.\label{completeLLL}
\end{align} 
%%%%%%%%%%%%%%%%%%%%%%%%%%%%%%%%%%%%%%%%%%%%%%%%%%%%%%%%%%%%%%%%%%%%%%%%%%
These states are holomorphic  
about $z$ and $\theta$, i.e. super-holomorphic except for their 
exponential term. 
 They have  angular momenta $m+1/2$ and $m+1$ respectively, and are  
localized on the same radially symmetric orbit with radius $R_{m+1}$. 
This reminds 
 the situation  where two particles, one of which
 has spin-0 and the other has spin-1/2 down, 
 rotate on a  plane with  the same radius [Fig.\ref{correspondingreal.fig}].
There appear no spin-1/2 up fermions in the LLL, and 
 the system shows  the super-chirality,
 where  not only the orbital rotations but also
  the spin rotations are chiral.
In the higher LLs, there are both spin-1/2 up and down 
 fermions, and the system is non-chiral. (See Fig.\ref{N2}.)

%%%%%%%%%%%%%%%%%%%%%%%%%%%%%%%%%%%%%%%%%%%%%%%%%%%%%%%%%%%%%%%%%5
%%%%%%%%%%%%%%%%%%%%%%%%%%%%%%%%%%%%%%%%%%%%%%%%%%%%%%%%%%%%%%%%%%

\section{Laughlin wavefunction and its superpartner}\label{Laughlinplane}

We construct a Laughlin wavefunction in the supersymmetric framework, 
by demanding  following  conditions 
as in the original case \cite{LaughlinPRL50}. 
The Laughlin wavefunction 
(i) is an eigenstate of $L_{\perp}$, (ii) possesses the  translational 
symmetries on the superplane up to its exponential factor.
We also postulate that the Laughlin wavefunction on the 
NC superplane is composed of a  
product of the bosonic part and the fermionic part.
It may be  natural to use   the 
original Laughlin wavefunction as the bosonic part. 
With respect to the fermionic part, the Vandermonde determinant vanishes
 due to the nilpotency of  the
 Grassmann  number, $\prod_{i<j}^{N}(\theta_i-\theta_j)=0$ for $N\ge 3$, 
and $(\theta_i-\theta_j)^m=0$ for $m\ge 2$.
Then, Laughlin wavefunction on the NC superplane is simply given by 
%%%%%%%%%%%%%%%%%%%%%%%%%%%%%%%%%%%%%%%%%%%%%%%%%%%%%%%%%%%%%%%%%%%%%%%%%%
\begin{align}
&\Psi_{Llin}=\prod_{p<q}^N(z_p-z_q)^m e^{-\sum_p(|z_p|^2+
(\theta\theta^*)_p)},
\label{superLlinplane}
\end{align}
%%%%%%%%%%%%%%%%%%%%%%%%%%%%%%%%%%%%%%%%%%%%%%%%%%%%%%%%%%%%%%%%%%%%%%%%%%
 where $N$ denotes the number of particles.
Apparently, $\Psi_{Llin}$ lives in the LLL, and 
is an eigenstate of $L_{\perp}$  with  eigenvalue $ mN(N-1)/2 $. 
Thus, $N$ particles described by $\Psi_{Llin}$ are spin-less particles,    
 which rotate on  the radially symmetric orbits in order from the origin. 
Intriguingly,  $\Psi_{Llin}$ has
  its superpartner $\Psi_{sLlin}$ unlike the 
 Laughlin-Haldane wavefunction on the  supersphere \cite{HasebePRL29}.
This stems from the decoupling between $X_i$ and  $\Theta_{\alpha}$ 
(\ref{xthetadecoup}) on the NC superplane.
The superpartner $\Psi_{sLin}$ is related to $\Psi_{Llin}$   by the 
non-dynamical supersymmetry, and is explicitly given by
%%%%%%%%%%%%%%%%%%%%%%%%%%%%%%%%%%%%%%%%%%%%%%%%%%%%%%%%%%%%%%%%%%%%%%%%%
\begin{equation}
\Psi_{sLlin}=\sum_{p<q}(\frac{\theta_p-\theta_q}{z_p-z_q})\cdot 
\Psi_{Llin},
\end{equation}
%%%%%%%%%%%%%%%%%%%%%%%%%%%%%%%%%%%%%%%%%%%%%%%%%%%%%%%%%%%%%%%%%%%%%%%%%%
which  has  the  angular momentum  $(mN(N-1)-1)/2$. 
It is noted that $\Psi_{sLlin}$ is not simply expressed as 
 a product of a bosonic part and a fermionic part.
The $N$-particle state 
 described by $\Psi_{sLlin}$ is  a  super-position of all possible 
states where the  $(N-1)$ spin-less particles and one spin-1/2 down
 particle rotate on the radially symmetric orbits in order from the origin. 
With the  definition of  the filling factor   $\nu\equiv \frac{N}{A/(2\pi\ell_B^2)}$
(where $A$ denotes the area on the superplane), 
$\Psi_{Llin}$ and $\Psi_{sLlin}$ may become two degenerate ground states 
of the supersymmetric QH systems at $\nu=1/m$, because they should have an identical energy due to the supersymmetry.

The density of $\Psi_{Llin}$ is   
%%%%%%%%%%%%%%%%%%%%%%%%%%%%%%%%%%%%%%%%%%%%%%%%%%%%%%%%%%%%%%%%%%%%%%%%%%
\begin{equation}
{\Psi_{Llin}}^{*}{\Psi_{Llin}}=e^{-\frac{2}{m}W},
\end{equation}
%%%%%%%%%%%%%%%%%%%%%%%%%%%%%%%%%%%%%%%%%%%%%%%%%%%%%%%%%%%%%%%%%%%%%%%%%%
where $W$ is interpreted as the supersymmetric extension of 
 the plasma potential,
%%%%%%%%%%%%%%%%%%%%%%%%%%%%%%%%%%%%%%%%%%%%%%%%%%%%%%%%%%%%%%%%%%%%%%%%%%
\begin{align}
&W=-\frac{m^2}{2}\sum_{p<q}
\ln|(x+iy)_p-(x+iy)_q|^2\nonumber\\
&~~~~~~~-\frac{mB}{4}\sum_p(|x+iy|^2+2\theta_1\theta_2)_p.
\end{align}
%%%%%%%%%%%%%%%%%%%%%%%%%%%%%%%%%%%%%%%%%%%%%%%%%%%%%%%%%%%%%%%%%%%%%%%%%%
The first term represents the interaction between particles with negative 
charge $m$ on the superplane.
The second term is interpreted as  a background   made by
 unit positive charged particles which are uniformly distributed 
 on the superplane with  density $\rho_{\Phi}=1/2\pi\ell_B^2$. 
This plasma analogy suggests that the state described by $\Psi_{Llin}$ 
becomes  energetically favorable at $\nu=1/m$, and  fundamental excitations 
carry a fractional charge $1/m$ as in the original case \cite{LaughlinPRL50}. 

%%%%%%%%%%%%%%%%%%%%%%%%%%%%%%%%%%%%%%%%%%%%%%%%%%%%%%%%%%%%%%%%%%%%%%%%%%
\section{Hall currents and Excited states}
\label{excitedplane}
%%%%%%%%%%%%%%%%%%%%%%%%%%%%%%%%%%%%%%%%%%%%%%%%%%%%%%%%%%%%%%%%%%%%%%%%%%

 The Hall currents on the superplane are expressed as 
%%%%%%%%%%%%%%%%%%%%%%%%%%%%%%%%%%%%%%%%%%%%%%%%%%%%%%%%%%%%%%%%%%%%%%%%%%
\begin{subequations}\begin{align}
&I_i=\frac{d}{dt}X_i=-i[X_i,V]=\epsilon_{ij}\ell_B^2 E_j,\\
&I_{\alpha}=\frac{d}{dt}\Theta_{\alpha}=-i[\Theta_{\alpha},V]=
i\ell_B^2(\sigma_3)_{\alpha\beta}E_{\beta},
\end{align}\end{subequations}
%%%%%%%%%%%%%%%%%%%%%%%%%%%%%%%%%%%%%%%%%%%%%%%%%%%%%%%%%%%%%%%%%%%%%%%%% 
where $\{E_i\}$ and $\{E_{\alpha}\}$ are bosonic and fermionic 
electric fields  defined by $E_i\equiv -\partial_i V$ and $E_{\alpha}
\equiv  -C_{\alpha\beta}\partial_{\beta}V$.
The Hall currents are orthogonal to the electric fields individually,
%%%%%%%%%%%%%%%%%%%%%%%%%%%%%%%%%%%%%%%%%%%%%%%%%%%%%%%%%%%%%%%%%%%%%%%%%%
\begin{equation}
E_i I_i=C_{\alpha\beta}E_{\alpha}I_{\beta}=0.
\end{equation}

\begin{figure}[tbph]
\includegraphics*[width=90mm]{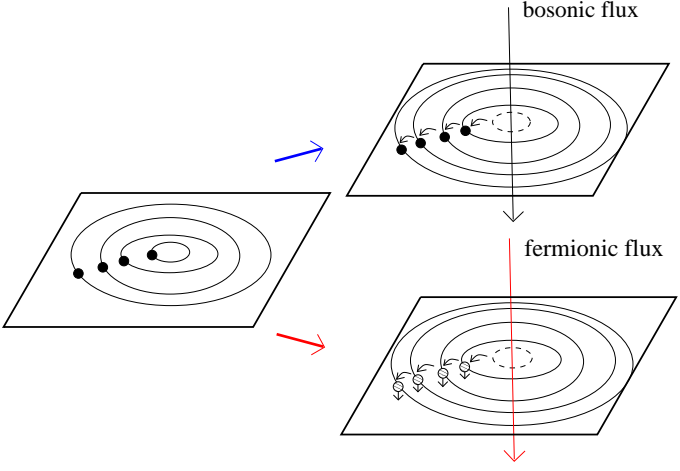}
\caption{In the left figure, the black blobs represent the spin-less 
particles described by  the Laughlin state for $\nu=1$. Due to  the 
  flux penetration, 
 the spin-less particles are pushed
 outwards   by $\delta R =\sqrt{2}\ell_B$, and 
  a quasi-hole is generated at the origin.
  The penetration of the bosonic flux keeps the particles spin-less, while
 the penetration of the fermionic flux changes the spin of each particle 
 from $0$ to $-1/2$. }
\label{pushedoutward.fig}
\vspace{-3mm}
\end{figure}

 As suggested by the existence of the 
 bosonic and fermionic Hall currents, 
there are two kinds of quasi-holes, one of which  is  
 bosonic 
  and the other is  fermionic. 
 They are superpartners, and are constructed by operating the creation operators  
%%%%%%%%%%%%%%%%%%%%%%%%%%%%%%%%%%%%%%%%%%%%%%%%%%%%%%%%%%%%%%%%%%%%%%%%%%
\begin{equation}
A^{\dagger}_{B}=\prod_p z_p,~~~
A^{\dagger}_F=\prod_p \theta_p,
\end{equation}
%%%%%%%%%%%%%%%%%%%%%%%%%%%%%%%%%%%%%%%%%%%%%%%%%%%%%%%%%%%%%%%%%%%%%%%%%%
 on the Laughlin wavefunction $\Psi_{Llin}$.  
They satisfy the commutation relations with the radius operator as 
%%%%%%%%%%%%%%%%%%%%%%%%%%%%%%%%%%%%%%%%%%%%%%%%%%%%%%%%%%%%%%%%%%%%%%%
\begin{equation}
[R^2, A^{\dagger}_B]=[R^2, A^{\dagger}_F]=2N\ell_B^2.
\end{equation}
%%%%%%%%%%%%%%%%%%%%%%%%%%%%%%%%%%%%%%%%%%%%%%%%%%%%%%%%%%%%%%%%%%%%%%%
These relations imply that  both $A_B^{\dagger}$ and $A_F^{\dagger}$ push
 each of the particles on the Laughlin state  outwards  
by $\delta R =\sqrt{2}\ell_B$, to generate 
a quasi-hole (or a new magnetic cell of the area $2\pi\ell_B^2$)
 at the origin.
Hence, 
the bosonic  and the fermionic quasi-holes carry the identical 
fractional charge $1/m$.
This may be regarded as  a consequence of supersymmetry, because superpartners
 should have same quantum numbers, such as mass, charge, except
 for  spin.
The commutation relations with the angular momentum 
are different 
%%%%%%%%%%%%%%%%%%%%%%%%%%%%%%%%%%%%%%%%%%%%%%%%%%%%%%%%%%%%%%%%%%%%%%%
\begin{equation}
[L_{\perp},A^{\dagger}_B]=N,~~[L_{\perp},A^{\dagger}_F]=\frac{N}{2},
\end{equation}
%%%%%%%%%%%%%%%%%%%%%%%%%%%%%%%%%%%%%%%%%%%%%%%%%%%%%%%%%%%%%%%%%%%%%%%
 which implies that, $A_B^{\dagger}$ does not change the spin of 
 each particle,
 while  $A_F^{\dagger}$   changes  the spin
  from $0$ to $-1/2$ [Fig.\ref{pushedoutward.fig}].

%%%%%%%%%%%%%%%%%%%%%%%%%%%%%%%%%%%%%%%%%%%%%%%%%%%%%%%%%%%%%%%%%5
%%%%%%%%%%%%%%%%%%%%%%%%%%%%%%%%%%%%%%%%%%%%%%%%%%%%%%%%%%%%%%%%%%

Similarly, bosonic  and  fermionic 
quasi-particle
 wavefunctions  would be  constructed by operating the annihilation operators 
%%%%%%%%%%%%%%%%%%%%%%%%%%%%%%%%%%%%%%%%%%%%%%%%%%%%%%%%%%%%%%%%%%%%%%%%%%
\begin{equation}
A_B=\prod_p\frac{\partial}{\partial z}_p,~~~
A_F=\prod_p\frac{\partial}{\partial \theta}_p,
\end{equation}
%%%%%%%%%%%%%%%%%%%%%%%%%%%%%%%%%%%%%%%%%%%%%%%%%%%%%%%%%%%%%%%%%%%%%%%%%%
%%%%%%%%%%%%%%%%%%%%%%%%%%%%%%%%%%%%%%%%%%%%%%%%%%%%%%%%%%%%%%%%%%%%%%%%%%
%%%%%%%%%%%%%%%%%%%%%%%%%%%%%%%%%%%%%%%%%%%%%%%%%%%%%%%%%%%%%%%%%%%%%%%%%%
on the Vandermonde determinant of $\Psi_{Llin}$.
$A_B$ and $A_F$ satisfy the commutation relations with the radius operator 
%%%%%%%%%%%%%%%%%%%%%%%%%%%%%%%%%%%%%%%%%%%%%%%%%%%%%%%%%%%%%%%%%%%%%%%
\begin{equation}
[R^2,A_B]=[R^2,A_F]=-2N\ell_B^2,
\end{equation}
%%%%%%%%%%%%%%%%%%%%%%%%%%%%%%%%%%%%%%%%%%%%%%%%%%%%%%%%%%%%%%%%%%%%%%%
and with the angular momentum  
%%%%%%%%%%%%%%%%%%%%%%%%%%%%%%%%%%%%%%%%%%%%%%%%%%%%%%%%%%%%%%%%%%%%%%%
\begin{equation}
[L_{\perp},A_B]=-N,~~~[L_{\perp},A_F]=-\frac{N}{2}.
\end{equation}
%%%%%%%%%%%%%%%%%%%%%%%%%%%%%%%%%%%%%%%%%%%%%%%%%%%%%%%%%%%%%%%%%%%%%%%
Thus,  $A_B$ attracts  each of the  particles on the Laughlin state
 by $\delta R =\sqrt{2}\ell_B$ 
inwards without changing its spin, 
and a bosonic quasi-particle with charge $-1/m$ is generated at the origin.
However, the operation of  $A_F$ on the 
 Vandermonde determinant of $\Psi_{Llin}$ yields zero, and fermionic quasi-particle excitations 
 do not appear in the LLL. % This stems from the super-chirality in the LLL. 
 It is because, while
 $A_F$ changes the spin of each particle from 0 to +1/2, 
 such spin-1/2 up particles are excluded due to the super-chiral property
 in the LLL.

$A_B$ and $A_B^{\dagger}$ satisfy the bosonic commutation relations,
%%%%%%%%%%%%%%%%%%%%%%%%%%%%%%%%%%%%%%%%%%%%%%%%%%%%%%%%%%%%%%%%%%%%%%%
\begin{subequations}
\begin{align}
&[A_B,A_B^{\dagger}]=1,\\
&[A_B,A_B]=[A_B^{\dagger},A_B^{\dagger}]=0,
\end{align}
\end{subequations}
%%%%%%%%%%%%%%%%%%%%%%%%%%%%%%%%%%%%%%%%%%%%%%%%%%%%%%%%%%%%%%%%%%%%%%%
while $A_F$ and $A_F^{\dagger}$ satisfy ``fermionic'' commutation
 relations 
%%%%%%%%%%%%%%%%%%%%%%%%%%%%%%%%%%%%%%%%%%%%%%%%%%%%%%%%%%%%%%%%%%%%%%%
\begin{subequations}
\begin{align}
&A_F A_F^{\dagger}+(-1)^N A_F^{\dagger}A_F=1,\\
&\{A_F,A_F\}=\{A_F^{\dagger},A_F^{\dagger}\}=0.
\end{align}
\end{subequations}
%%%%%%%%%%%%%%%%%%%%%%%%%%%%%%%%%%%%%%%%%%%%%%%%%%%%%%%%%%%%%%%%%%%%%%%
%Thus, when $N$ is even, $A_F$ and $A_F^{\dagger}$ become  hard-core
% bosonic operators.

\section{Relations to  bilayer QH systems}\label{realsystem}

 It is well known,  fermionic harmonic oscillators 
 can be  regarded as the spin-1/2 ladder operators in supersymmetric quantum
 mechanics. 
In fact, 
the ladder operators made by Pauli matrices, $(\sigma_+,\sigma_-)=
\frac{1}{2}(\sigma_1+i\sigma_2,\sigma_1-i\sigma_2)$, satisfy the 
equations
%%%%%%%%%%%%%%%%%%%%%%%%%%%%%%%%%%%%%%%%%%%%%%%%%%%%%%%%%%%%%%%%%%%%%%%%%%
\begin{equation}
\{\sigma_+,\sigma_-\}=1,~~\sigma_+^2=\sigma_-^2=1,
\end{equation}
%%%%%%%%%%%%%%%%%%%%%%%%%%%%%%%%%%%%%%%%%%%%%%%%%%%%%%%%%%%%%%%%%%%%%%%%%
 which are equivalent to the  properties of the fermionic harmonic 
oscillators.
Due to this identification, it is possible to map 
  a supersymmetric harmonic oscillator system to a spin system.
 In the
 supersymmetric QH system,
 there exist two kinds of fermionic harmonic oscillators,
 $(\alpha,\alpha^{\dagger})$ and $(\beta,\beta^{\dagger})$. 
Therefore, in its corresponding spin system, two kinds of ``spins'' 
are needed.
One possible candidate to meet this requirement is a bilayer QH
 system, where electrons carry  
 not only their intrinsic spins  but also 
pseudospins which specify double layers.
By regarding $\alpha$-``spin'' as pseudospin  and $\beta$-``spin'' as
 intrinsic spin, there exists a  mapping 
 to bilayer QH systems [Table \ref{correspondence}].
 However, unfortunately, the real boson-fermion symmetry 
 in the supersymmetric QH system is lost 
 in this mapping,
 since the corresponding
 $\mathcal{N}=1$ non-dynamical supersymmetry  in the bilayer 
 QH system act as interchange of  
 the spin-1/2 up and down fermions.

When, we assign $\alpha$-``spin'' as 
%%%%%%%%%%%%%%%%%%%%%%%%%%%%%%%%%%%%%%%%%%%%%%%%%%%%%%%%%%%%%%%%%%%%%%%%
\begin{equation}
%$
(\alpha,\alpha^{\dagger})\leftrightarrow (\tau_+,\tau_-)\equiv 
\frac{1}{2}(\tau_2+i\tau_3,\tau_2-i\tau_3),
\end{equation}
%%%%%%%%%%%%%%%%%%%%%%%%%%%%%%%%%%%%%%%%%%%%%%%%%%%%%%%%%%%%%%%%%%%%%%%%
%%%%%%%%%%%%%%%%%%%%%%%%%%%%%%%%%%%%%%%%%%%%%%%%%%%%%%%%%%%%%%%%%%%%%%%
where $\{\tau_a\} (a=1,2,3)$  represent Pauli matrices for the 
pseudospin,
the Hamiltonian (\ref{hamilsecond}) is  rewritten as 
%%%%%%%%%%%%%%%%%%%%%%%%%%%%%%%%%%%%%%%%%%%%%%%%%%%%%%%%%%%%%%%%%%%%%%%
\begin{equation}
H={\omega}(a^{\dagger}a+\frac{1}{2})-\frac{\omega}{2}\tau_1,
\label{bilayerlimithamil}
\end{equation}
%%%%%%%%%%%%%%%%%%%%%%%%%%%%%%%%%%%%%%%%%%%%%%%%%%%%%%%%%%%%%%%%%%%%%%%
which is the non-Coulomb part of the 
 Hamiltonian for  bilayer QH systems,
 with tunneling interaction 
$\Delta_{SAS}=\omega$ and   without Zeeman interaction $\Delta_Z=0$. 
The LLL in supersymmetric QH systems can be  regarded as  
 the LLL of symmetric layer state 
in  bilayer QH systems.
%%%%%%%%%%%%%%%%%%%%%%%%%%%%%%%%%%%%%%%%%%%%%%%%%%%%%%%%%%%%%%%%%%%%
%%%%%%%%%%%%%%%%%%%%%
\begin{table}
\renewcommand{\arraystretch}{1.5}
\begin{center}
\begin{tabular}{|c||c|}  
  \hline Supersymmetric QH system &  Bilayer QH  system \\  
\hline \hline Bosonic oscillator $a$  & Landau levels     
\\ \hline  
 Fermionic oscillator $\alpha$ &  Bilayers or pseudospins
\\ \hline    
Bosonic oscillator  $b$ & Radially symmetric orbits \\  \hline
 Fermionic oscillator $\beta$ &  (Intrinsic) spins
\\ \hline    
\end{tabular}
\end{center}
\caption{The supersymmetric 
 QH system is mapped to a bilayer QH system.
The fermionic operators $(\alpha,\alpha^{\dagger})$ 
and $(\beta,\beta^{\dagger})$ are regarded as the ladder operators 
 for pseudospin and intrinsic spin.
The bosonic operators $(a,a^{\dagger})$ and $(b,b^{\dagger})$ 
are identified with the ladder operators for LLs and 
 radially symmetric orbits. } 
\label{correspondence}
\end{table}
%%%%%%%%%%%%%%%%%%%%%%%%%%%%%%%%%%%%%%%%%%%%%%%%%%%%%%%%%%%%%%%%%%%%%%
The Hamiltonian 
(\ref{bilayerlimithamil})
 appears in many different context 
 of supersymmetric quantum mechanical systems, such as 
 Pauli Hamiltonian with  gyromagnetic factor 2 \cite{Crombrugghe1983} and 
the Jaynes-Cummings model without interaction terms used in quantum 
optics \cite{SuperJCM}.
 However, it must be noted that each of such systems possesses  
$\mathcal{N}=1$ supersymmetry, while the present QH system has 
 larger $\mathcal{N}=2$ supersymmetry due to the existence of extra  
$\mathcal{N}$=1 non-dynamical supersymmetry. 
\section{summary and discussion}\label{summary}

%%%%%%%%%%%%%%%%%%%%%%%%%%%%%%%%%%%%%%%%%%%%%%%%%%%%%%%%%%%%%%%%%%%%%%%%%%
%%%%%%%%%%%%%%%%%%%%%%%%%%%%%%%%%%%%%%%%%%%%%%%%%%%%%%%%%%%%%%%%%%%%%%%%%%

Based on the  supersymmetric NC algebra, 
we  constructed  QH liquids on a NC superplane. 
The supersymmetric Landau model enjoys (complex) $\mathcal{N}=2$ supersymmetry, one of which is dynamical and the other is non-dynamical.
In the LLL, only the $\mathcal{N}=1$ non-dynamical supersymmetry remains valid.
Unlike ordinary supersymmetric quantum mechanics,  
 the present supersymmetry  represents a real boson-fermion symmetry.
The NAC fermionic coordinates are 
related to   spin degrees of freedom, and bring the 
super-chiral property to the LLL. 
Since, on the NC superplane, the bosonic 
 and the fermionic center-of-mass coordinates are 
decoupled, the Laughlin wavefunction and topological 
excitations  have their superpartners unlike the QH liquid on 
the fuzzy supersphere.
With use of the identification between 
 the fermionic harmonic operators and 
 the ``spin''-1/2
 ladder operators, supersymmetric QH systems are mapped to  
 bilayer QH systems.
 In this mapping,
 the LLL in  supersymmetric QH systems is regarded as the 
 LLL in the symmetric layer state of  bilayer QH systems.

 While we have clarified  bulk properties in the supersymmetric QH liquid, 
 it is also important  to study 
  its edge excitations and 
 effective field theory for  further understanding 
 of  physics of  the NAC geometry.
 We would like to pursue them in a future publication.

%%%%%%%%%%%%%%%%%%%%%%%%%%%%%%%%%%%%%%%%%%%%%%%%%%%%%%%%%%%%%%%%%%%%%%%%%%%
\section*{ACKNOWLEDGEMENTS}
 I would like   to acknowledge  Satoshi Iso and
 Hiroshi Umetsu  for useful discussions.
 I also thank  Bernard deWitt, Shinsuke Kawai and Giovanni Landi for 
helpful conversations.
 This work was supported by a JSPS fellowship. 
%%%%%%%%%%%%%%%%%%%%%%%%%%%%%%%%%%%%%%%%%%%%%%%%%%%%%%%%%%%%%%%%%%%%%%%%%%
\appendix
%%%%%%%%%%%%%%%%%%%%%%%%%%%%%%%%%%%%%%%%%%%%%%%%%%%%%%%%%%%%%%%%%%%%%%%%%%

%%%%%%%%%%%%%%%%%%%%%%%%%%%%%%%%%%%%%%%%%%%%%%%%%%%%%%%%%%%%%%%%%%%%%%%
%%%%%%
\section{Magnetic translations on the superplane}\label{magnetictrans}
%%%%%%%%%%%%%%%%%%%%%%%%%%%%%%%%%%%%%%%%%%%%%%%%%%%%%%%%%%%%%%%%%%%%%%%%%%

 In this section, we summarize  
Aharonov-Bohm phase accompanied by  magnetic translation on the 
 NC superplane.
With use of the center-of-mass coordinates $(X_i,\Theta_{\alpha})$, 
the supersymmetric magnetic translation operator is constructed  as 
%%%%%%%%%%%%%%%%%%%%%%%%%%%%%%%%%%%%%%%%%%%%%%%%%%%%%%%%%%%%%%%%%%%%%%%%%
\begin{equation}
\mathcal{T}_K=e^{i(k_iX_i+\kappa_{\alpha}\Theta_{\alpha})},
\end{equation}
%%%%%%%%%%%%%%%%%%%%%%%%%%%%%%%%%%%%%%%%%%%%%%%%%%%%%%%%%%%%%%%%%%%%%%%%%%
which satisfies 
%%%%%%%%%%%%%%%%%%%%%%%%%%%%%%%%%%%%%%%%%%%%%%%%%%%%%%%%%%%%%%%%%%%%%%%%%
\begin{equation}
\mathcal{T}_K\cdot \mathcal{T}_T=\mathcal{T}_{K+T}e^{-\frac{1}{2}
\ell_B^2\Sigma_{IJ}K_I T_J},
\end{equation}
%%%%%%%%%%%%%%%%%%%%%%%%%%%%%%%%%%%%%%%%%%%%%%%%%%%%%%%%%%%%%%%%%%%%%%%%%
where  $K\equiv (k_i,\kappa_{\alpha})$, $T\equiv (t_i,\tau_{\alpha})$ and  
%%%%%%%%%%%%%%%%%%%%%%%%%%%%%%%%%%%%%%%%%%%%%%%%%%%%%%%%%%%%%%%%%%%%%%%%%
%\begin{equation}
$\Sigma\equiv 
\begin{pmatrix}
&\sigma_2 & 0\\ 
&0 & -\sigma_1 
\end{pmatrix}$.
%\end{equation}
%%%%%%%%%%%%%%%%%%%%%%%%%%%%%%%%%%%%%%%%%%%%%%%%%%%%%%%%%%%%%%%%%%%%%%%%%
The  algebra for the super-magnetic translation is given by 
%%%%%%%%%%%%%%%%%%%%%%%%%%%%%%%%%%%%%%%%%%%%%%%%%%%%%%%%%%%%%%%%%%%%%%%%%
\begin{align}
&[\mathcal{T}_K,\mathcal{T}_T]=-2\mathcal{T}_{K+T}\cdot  
\sinh\biggl(\frac{1}{2}\Sigma_{IJ}K_I T_J\biggr)
 \nonumber\\
&=2\mathcal{T}_{K+T}\cdot [e^{-\frac{1}{2}\ell_B^2 
(\sigma_2)_{ij}k_it_j}\sinh (\frac{1}{2}\ell_B^2
(\sigma_1)_{\alpha\beta}\kappa_{\alpha}\tau_{\beta})\nonumber\\
&~~~~~~~~~~~~~~~~~ +ie^{-\frac{1}{2}\ell_B^2(\sigma_1)_{\alpha\beta}
\kappa_{\alpha}\tau_{\beta}}\sin(\frac{1}{2}\ell_B^2\epsilon_{ij}k_it_j)].
\end{align}
%%%%%%%%%%%%%%%%%%%%%%%%%%%%%%%%%%%%%%%%%%%%%%%%%%%%%%%%%%%%%%%%%%%%%%%%%
%In the first equation, we have used $\Sigma_{IJ}K_IT_J=-
%\Sigma_{IJ}T_IK_J$. 
The  round-trip acquires a supersymmetric Aharonov-Bohm phase as
%%%%%%%%%%%%%%%%%%%%%%%%%%%%%%%%%%%%%%%%%%%%%%%%%%%%%%%%%%%%%%%%%%%%%%%%%
\begin{equation}
\mathcal{T}_{-K}\mathcal{T}_{-T}\mathcal{T}_{K}\mathcal{T}_{T}=e^{-BS},
\end{equation}
%%%%%%%%%%%%%%%%%%%%%%%%%%%%%%%%%%%%%%%%%%%%%%%%%%%%%%%%%%%%%%%%%%%%%%%%%%
where 
%%%%%%%%%%%%%%%%%%%%%%%%%%%%%%%%%%%%%%%%%%%%%%%%%%%%%%%%%%%%%%%%%%%%%%%%%%
\begin{equation}
S\equiv \ell_B^4\Sigma_{IJ}K_I T_J=\ell_B^4((\sigma_2)_{ij}k_ip_j
-(\sigma_1)_{\alpha\beta}\kappa_{\alpha}\tau_{\beta}),
\end{equation}
%%%%%%%%%%%%%%%%%%%%%%%%%%%%%%%%%%%%%%%%%%%%%%%%%%%%%%%%%%%%%%%%%%%%%%%%%
 which represents the ``area'' on the superplane.

%%%%%%%%%%%%%%%%%%%%%%%%%%%%%%%%%%%%%%%%%%%%%%%%%%%%%%%%%%%%%%%%%%%%%%%%%
%%%%%%%%%%%%%%%%%%%%%%%%%%%%%%%%%%%%%%%%%%%%%%%%%%%%%%%%%%%%%%%%%%%%%%%
\section{Infinite Symmetries in the LLL }\label{infinitsym}

It is well known, in the LLL, infinite conserved charges appear and 
form  the $W_{\infty}$ algebra \cite{hep-th/9206027,IsoPLB92}.
Similarly, a  supersymmetric extension of the $W_{\infty}$ algebra appears
in the LLL of the supersymmetric QH systems.
 It is obvious the following quantities commute with the Hamiltonian (\ref{hamilsecond}),
%%%%%%%%%%%%%%%%%%%%%%%%%%%%%%%%%%%%%%%%%%%%%%%%%%%%%%%%%%%%%%%%%%%%%%%%%%
\begin{subequations}
\begin{align}
&L^B_{m,n}=(b^{\dagger})^{m+1} b^{n+1},\\
& {L}^F_{m,n}=(b^{\dagger})^{m+1} b^{n+1}\beta, \\
&{{L}^{F}}_{m,n}^{\dagger}  =(b^{\dagger})^{n+1} b^{m+1}\beta^{\dagger}, 
\end{align}
\end{subequations}
%%%%%%%%%%%%%%%%%%%%%%%%%%%%%%%%%%%%%%%%%%%%%%%%%%%%%%%%%%%%%%%%%%%%%%%%%%
and ${L^B}_{m,n}^{\dagger}=L^B_{n,m}$, where $m,n\ge -1$.
In particular, non-dynamical supercharges are identified as 
$(\tilde{Q},\tilde{Q}^{\dagger})
=(L_{0,-1}^F,{L^F}^{\dagger}_{0,-1} )$. 
$L^B_{m,n}$ and $L^F_{k,l}$ satisfy a supersymmetric extension 
 of the $W_{\infty}$  algebra as 
%%%%%%%%%%%%%%%%%%%%%%%%%%%%%%%%%%%%%%%%%%%%%%%%%%%%%%%%%%%%%%%%%%%%%%%%%
\begin{subequations}
\begin{align}
&[L^B_{m,n},L^B_{k,l}]\nonumber\\
&~~=
\sum_{s=0}^{Min(n,k)}\frac{(n+1)!(k+1)!}{(n-s)!(k-s)!(s+1)!}
L^B_{m+k-s, n+l-s}\nonumber\\
&~~~~~~~-    ((m,n)\leftrightarrow (k,l)),\label{superW1} \\
&[L^B_{m,n},{L}^F_{k,l}]\nonumber\\
&~~=\sum_{s=0}^{Min(n,k)}\frac{(n+1)!(k+1)!}{(n-s)!(k-s)!(s+1)!}
L^F_{m+k-s, n+l-s}\nonumber\\
&~~~~~~~-    ((m,n)\leftrightarrow (k,l)), \\
&\{{L}^F_{m,n},{L}^F_{k,l}\}=0.
\end{align}\label{superWalgebra1}
\end{subequations}
%%%%%%%%%%%%%%%%%%%%%%%%%%%%%%%%%%%%%%%%%%%%%%%%%%%%%%%%%%%%%%%%%%%%%%%%%%
$L^B_{m,n}$ and ${{L^F}_{k,l}}^{\dagger} $  satisfy another    
  supersymmetric  $W_{\infty}$ algebra similar to Eq.(\ref{superWalgebra1}).
The commutation relations with 
 the angular momentum and the radius operator are given by 
%%%%%%%%%%%%%%%%%%%%%%%%%%%%%%%%%%%%%%%%%%%%%%%%%%%%%%%%%%%%%%%%%%%%%%
\begin{subequations}
\begin{align}
&[L_{\perp},L_{m,n}^B]=(m-n)L_{m,n}^B,\\
&[L_{\perp},L_{m,n}^F]=(m-n-\frac{1}{2})L_{m,n}^F,\\
&[L_{\perp},{L^F}^{\dagger}_{m,n} ]=(-m+n+\frac{1}{2})
{L^F}^{\dagger}_{m,n},
\end{align}\label{commuangL}
\end{subequations}
%%%%%%%%%%%%%%%%%%%%%%%%%%%%%%%%%%%%%%%%%%%%%%%%%%%%%%%%%%%%%%%%%%%%%%%
and 
%%%%%%%%%%%%%%%%%%%%%%%%%%%%%%%%%%%%%%%%%%%%%%%%%%%%%%%%%%%%%%%%%%%%%%
\begin{subequations}
\begin{align}
&[R^2,L_{m,n}^B]=(m-n)L_{m,n}^B,\\
&[R^2,L_{m,n}^F]=(m-n-{1})L_{m,n}^F,\\
&[R^2,{L^F}^{\dagger}_{m,n}]=(-m+n+{1}){L^F}^{\dagger}_{m,n} .
\end{align}\label{commuradL}
\end{subequations}
%%%%%%%%%%%%%%%%%%%%%%%%%%%%%%%%%%%%%%%%%%%%%%%%%%%%%%%%%%%%%%%%%%%%%%%
These relations imply that radially symmetric orbits (\ref{otherbasisLLL}) are 
related by $L_{m,n}^B$, $L_{m,n}^F$ and 
${L^F}^{\dagger}_{m,n}$ as 
%%%%%%%%%%%%%%%%%%%%%%%%%%%%%%%%%%%%%%%%%%%%%%%%%%%%%%%%%%%%%%%%%%%%%%%
\begin{subequations}
\begin{align}
&|m+1/2>=\frac{1}{\sqrt{m!n!}}L_{m-1,n-1}^B|n+1/2>,\\
&|m+1>=\frac{1}{\sqrt{(m+1)!(n+1)!}}L_{m,n}^B|n+1>,
\end{align}
\end{subequations}
%%%%%%%%%%%%%%%%%%%%%%%%%%%%%%%%%%%%%%%%%%%%%%%%%%%%%%%%%%%%%%%%%%%%%%
and 
%%%%%%%%%%%%%%%%%%%%%%%%%%%%%%%%%%%%%%%%%%%%%%%%%%%%%%%%%%%%%%%%%%%%%%%
\begin{subequations}
\begin{align}
&|m+1/2>=\frac{1}{\sqrt{m!(n+1)!}}{L^F}^{\dagger}_{n,m-1}|n+1>,\\
&|m+1>=\frac{1}{\sqrt{(m+1)!n!}}L_{m,n-1}^F|n+1/2>.
\end{align}
\end{subequations}
%%%%%%%%%%%%%%%%%%%%%%%%%%%%%%%%%%%%%%%%%%%%%%%%%%%%%%%%%%%%%%%%%%%%%%

\section{Radially Symmetric Orbits}\label{radiallysymorbits}

Since the Hamiltonian for the supersymmetric Landau problem 
 (\ref{hamilsecond})
is given by a  sum 
of the  bosonic oscillators and the fermionic oscillators, 
 the whole supersymmetric Hilbert space 
is simply constructed by a direct product of bosonic and fermionic 
Hilbert spaces.
In this section, with use of the symmetric gauge, we present
 explict forms of the basis in bosonic and fermionic Landau problems.

%\subsection{The bosonic Landau problem}

First, we concisely review the bosonic Landau problem.
The Hamiltonian and the angular momentum are given by 
$H_{B}=\omega(a^{\dagger}a+1/2)$ and $L_B=b^{\dagger}b-a^{\dagger}a$,
 respectively. 
The state in the bosonic Hilbert space with energy 
$E_n=\omega(n+{1}/{2})$ and  angular momentum $l=m-n$ is  
%%%%%%%%%%%%%%%%%%%%%%%%%%%%%%%%%%%%%%%%%%%%%%%%%%%%%%%%%%%%%%%%%%%%%%%%%%
\begin{equation}
|n,l>=\sqrt{\frac{1}{n!m!}}(a^{\dagger})^n(b^{\dagger})^{m}|0>.
\end{equation}
%%%%%%%%%%%%%%%%%%%%%%%%%%%%%%%%%%%%%%%%%%%%%%%%%%%%%%%%%%%%%%%%%%%%%%%%%%
In particular, the Hilbert space in  LLL ($n=0$) is spanned by the basis 
%%%%%%%%%%%%%%%%%%%%%%%%%%%%%%%%%%%%%%%%%%%%%%%%%%%%%%%%%%%%%%%%%%%%%%%%%%
\begin{equation} 
|m>=\frac{1}{\sqrt{m!}}(b^{\dagger})^m|0>.
\label{LLLbasis1}
\end{equation}
%%%%%%%%%%%%%%%%%%%%%%%%%%%%%%%%%%%%%%%%%%%%%%%%%%%%%%%%%%%%%%%%%%%%%%%%%
When we adopt the symmetric gauge, the LLL condition, 
$a|\text{LLL}\!>=0$, is  denoted as   
%%%%%%%%%%%%%%%%%%%%%%%%%%%%%%%%%%%%%%%%%%%%%%%%%%%%%%%%%%%%%%%%%%%%%%%%%
\begin{equation}
(z+\partial^*)\phi_{LLL}=0.
\end{equation}
%%%%%%%%%%%%%%%%%%%%%%%%%%%%%%%%%%%%%%%%%%%%%%%%%%%%%%%%%%%%%%%%%%%%%%%%%
Hence, the wavefunction in LLL is generally expressed  as 
%%%%%%%%%%%%%%%%%%%%%%%%%%%%%%%%%%%%%%%%%%%%%%%%%%%%%%%%%%%%%%%%%%%%%%%%%%
\begin{equation} 
\phi_{LLL}=f(z)e^{-|z|^2},
\end{equation}
%%%%%%%%%%%%%%%%%%%%%%%%%%%%%%%%%%%%%%%%%%%%%%%%%%%%%%%%%%%%%%%%%%%%%%%%%
where $f(z)$ is an arbitrary holomorphic function and 
  any wavefunction in  LLL can be expanded by  the radially symmetric  
orbits, 
%%%%%%%%%%%%%%%%%%%%%%%%%%%%%%%%%%%%%%%%%%%%%%%%%%%%%%%%%%%%%%%%%%%%%%%%%%
\begin{equation}
\phi_m=\sqrt{\frac{2^{m+1}}{\pi m!}}z^m e^{-|z|^2}.
\end{equation}
%%%%%%%%%%%%%%%%%%%%%%%%%%%%%%%%%%%%%%%%%%%%%%%%%%%%%%%%%%%%%%%%%%%%%%%%%%
They are the position representation of Eq.(\ref{LLLbasis1}) and satisfy 
the orthonormal condition
%%%%%%%%%%%%%%%%%%%%%%%%%%%%%%%%%%%%%%%%%%%%%%%%%%%%%%%%%%%%%%%%%%%%%%%%%
\begin{equation}
\int dz dz^* \phi_m^*(z,z^*) \phi_{m'} (z,z^*)=\delta_{mm'}.
\end{equation}
%%%%%%%%%%%%%%%%%%%%%%%%%%%%%%%%%%%%%%%%%%%%%%%%%%%%%%%%%%%%%%%%%%%%%%%%%
The ``complete relation'' in LLL is calculated as 
%%%%%%%%%%%%%%%%%%%%%%%%%%%%%%%%%%%%%%%%%%%%%%%%%%%%%%%%%%%%%%%%%%%%%%%%%
\begin{equation}
\sum_{m=0}^{\infty} \phi_m(z',z'^*){\phi_m}^*(z,z^*)
=\frac{2}{\pi}e^{-|z|^2-|z'|^2-2z'^*z}.\label{bosoncompleteLLL}
\end{equation}
%%%%%%%%%%%%%%%%%%%%%%%%%%%%%%%%%%%%%%%%%%%%%%%%%%%%%%%%%%%%%%%%%%%%%%%%%

%\subsection{The fermionic Landau problem}

The fermionic Landau problem is similarly analyzed.
The Hamiltonian and the angular momentum are given by 
$H_{F}=\omega(\alpha^{\dagger}\alpha-1/2)$ and 
$L_F=1/2(\beta^{\dagger}\beta-\alpha^{\dagger}\alpha)$,
 respectively. 
Due to the Pauli exclusion principle, 
the  Hilbert space for fermionic oscillators consists of  only 
four states.
There are only two energy levels, LLL and 1-st LL, with  energy
 $-\omega/2,~ \omega/2$, both of which are doubly degenerate. 
Two states in the  LLL with  angular momentum $0,~1/2$ are 
 given by
%%%%%%%%%%%%%%%%%%%%%%%%%%%%%%%%%%%%%%%%%%%%%%%%%%%%%%%%%%%%%%%
%%%%%%%%%%%
\begin{equation}
|0,0>=|0>,~~|0,1/2>=\beta^{\dagger}|0>,
\end{equation}
%%%%%%%%%%%%%%%%%%%%%%%%%%%%%%%%%%%%%%%%%%%%%%%%%%%%%%%%%%%%%
%%%%%%%%%%%%%
where $|0>$ is defined as $\alpha|0>=\beta|0>=0$.
Two states in  the 1-st LL  with  angular momentum $-1/2,~0$
 are given by
%%%%%%%%%%%%%%%%%%%%%%%%%%%%%%%%%%%%%%%%%%%%%%%%%%%%%%%%%%%%%%%
%%%%%%%%%%%
\begin{equation}
|1,-1/2> =\alpha^{\dagger}|0>, ~~ |1,0>=\alpha^{\dagger}
\beta^{\dagger}|0>.
\end{equation}
%%%%%%%%%%%%%%%%%%%%%%%%%%%%%%%%%%%%%%%%%%%%%%%%%%%%%%%%%%%%%%%
%%%%%%%%%%% 
In the symmetric gauge, the LLL condition, 
$\alpha|\text{LLL}\!>=0$, is  
rewritten as 
%%%%%%%%%%%%%%%%%%%%%%%%%%%%%%%%%%%%%%%%%%%%%%%%%%%%%%%%%%%%%%%
%%%%%%%%%%
\begin{equation}
(\theta-\partial_{\theta}^*)\varphi_{LLL}=0.
\end{equation}
%%%%%%%%%%%%%%%%%%%%%%%%%%%%%%%%%%%%%%%%%%%%%%%%%%%%%%%%%%%%%%%
%%%%%%%%%%
Hence, the wavefunction in fermionic LLL is generally  given 
by 
%%%%%%%%%%%%%%%%%%%%%%%%%%%%%%%%%%%%%%%%%%%%%%%%%%%%%%%%%%%%%%%%%
%%%%%%%%%
\begin{equation} 
\varphi_{LLL}=g(\theta)e^{-\theta\theta^*},
\end{equation}
%%%%%%%%%%%%%%%%%%%%%%%%%%%%%%%%%%%%%%%%%%%%%%%%%%%%%%%%%%%%%%%%
%%%%%%%%%
where $g(\theta)=g_0+g_1\theta$ is an arbitrary holomorphic 
function.
Therefore,  any wavefunction in the LLL of the fermionic oscillators
 can 
be expanded by  the following  states
%%%%%%%%%%%%%%%%%%%%%%%%%%%%%%%%%%%%%%%%%%%%%%%%%%%%%%%%%%%%%%%%%%%
%%%%%%%
\begin{subequations}
\begin{align}
&\varphi_{0,0}=\frac{1}{\sqrt{2}}e^{-\theta\theta^*}
=\frac{1}{\sqrt{2}}(1-\theta\theta^*), \\
&\varphi_{0,1/2}= \theta e^{-\theta\theta^*} = \theta.
\end{align}
\end{subequations}
%%%%%%%%%%%%%%%%%%%%%%%%%%%%%%%%%%%%%%%%%%%%%%%%%%%%%%%%%%%%%%%%%%%
%%%%%%%
In fact, these are the position representation of $|0,0>$ and 
$|0,1/2>$.
Similarly the position representation of the 1-st LL states,  
$|1,-1/2>$ and $|1,0>$, are 
%%%%%%%%%%%%%%%%%%%%%%%%%%%%%%%%%%%%%%%%%%%%%%%%%%%%%%%%%%%%%%%%%%%
%%%%%%
\begin{subequations}
\begin{align}
&\varphi_{1,-1/2}=\theta^*e^{\theta\theta^*} =\theta^*,\\    
&\varphi_{1,0}   = \frac{1}{\sqrt{2}}e^{\theta\theta^*}
=\frac{1}{\sqrt{2}}(1+\theta\theta^*).
\end{align}
\end{subequations}
%%%%%%%%%%%%%%%%%%%%%%%%%%%%%%%%%%%%%%%%%%%%%%%%%%%%%%%%%%%%%%%%%%
%%%%%%
They satisfy the orthonormal condition
%%%%%%%%%%%%%%%%%%%%%%%%%%%%%%%%%%%%%%%%%%%%%%%%%%%%%%%%%%%%%%%%%%%
%%%%%%
\begin{equation}
\int d\theta d\theta^* (-1)^{(n+1)} \varphi^*_{n,l} (\theta,\theta^*) 
\varphi_{n',l'}(\theta,\theta^*)= \delta_{n,n'}\delta_{l,l'},
\label{fermiortho}
\end{equation}
%%%%%%%%%%%%%%%%%%%%%%%%%%%%%%%%%%%%%%%%%%%%%%%%%%%%%%%%%%%%%%%%%%%%%%%%%
where we have included a weight factor and have defined the Grassmann 
integral as $\int d\theta d\theta^*\equiv
\partial_{\theta^*}\partial_{\theta}$.
The complete relation for these states is obtained as 
%%%%%%%%%%%%%%%%%%%%%%%%%%%%%%%%%%%%%%%%%%%%%%%%%%%%%%%%%%%%%%%%%%%%%%%%%
\begin{equation}
\sum_{n,l}(-1)^{n+1}\varphi_{n,l}(\theta,\theta^*)\varphi_{n,l}^*
(\theta',\theta'^{*})
=\delta(\theta-\theta')\delta(\theta^*-\theta'^*),
\end{equation} 
%%%%%%%%%%%%%%%%%%%%%%%%%%%%%%%%%%%%%%%%%%%%%%%%%%%%%%%%%%%%%%%%%%%%%%%%%%
where we have  taken into account  the weight factor 
as in Eq.(\ref{fermiortho}).
The ``complete  relation'' in the fermionic LLL is calculated as 
%%%%%%%%%%%%%%%%%%%%%%%%%%%%%%%%%%%%%%%%%%%%%%%%%%%%%%%%%%%%%%%%%%%%%%%%%
\begin{equation}
\sum_{l=0,1/2} \varphi_{0,l}(\theta,\theta^*){\varphi_{0,l}}^*
(\theta',\theta'^*)=\frac{1}{2}e^{-\theta\theta^*-\theta'\theta'^*-
2\theta'^*\theta}.\label{fermioncompleteLLL}
\end{equation}
%%%%%%%%%%%%%%%%%%%%%%%%%%%%%%%%%%%%%%%%%%%%%%%%%%%%%%%%%%%%%%%%%%%%%%%%%

%%%%%%%%%%%%%%%%%%%%%%%%%%%%%%%%%%%%%%%%%%%%%%%%%%%%%%%%%%%%%%%%%%%%%%%
%%%%%%%
\section{von Neumann basis on the superplane}\label{Neumann}
%%%%%%%%%%%%%%%%%%%%%%%%%%%%%%%%%%%%%%%%%%%%%%%%%%%%%%%%%%%%%%%%%%%%%%%
%%%%%%%%%%%%%%%%%%%%%%%%%%%%%%%%%%%%%%%%%%%%%%%%%%%%%%%%%%%%%%%%%%%%%%%
%%%%%%

In this section, we briefly discuss  von Neumann basis formalism on
  the superplane.
The von Neumann basis is 
a  convenient basis to investigate QH systems, because it is a 
 quantum mechanical analogue of
 the classical cyclotron orbit and, in a continuum limit, the translational 
symmetries are expected to be recovered \cite{ImaiPRB42}.
% The von Neumann basis consists of a subset of  coherent states, and spans
% the Hilbert space in the LLL.

First, we introduce  
the supercoherent state  as a simultaneous 
eigenstate  
of  two annihilation operators $\hat{b}$ and $\hat{\beta}$,  
%%%%%%%%%%%%%%%%%%%%%%%%%%%%%%%%%%%%%%%%%%%%%%%%%%%%%%%%%%%%%%%%%%%%%%%%%%
\begin{equation}
(\hat{b}+\hat{\beta})|b,\beta>=(b+\beta)|b,\beta>.
\end{equation}
%%%%%%%%%%%%%%%%%%%%%%%%%%%%%%%%%%%%%%%%%%%%%%%%%%%%%%%%%%%%%%%%%%%%%%%%%%
(The annihilation operators are denoted with hat to
 distinguish their eigenvalues.)
Explicitly, the supercoherent state is given by   
%%%%%%%%%%%%%%%%%%%%%%%%%%%%%%%%%%%%%%%%%%%%%%%%%%%%%%%%%%%%%%%%%%%%%%%%%%
\begin{equation}
|b,\beta>=|b>\otimes|\beta>,
\end{equation}
%%%%%%%%%%%%%%%%%%%%%%%%%%%%%%%%%%%%%%%%%%%%%%%%%%%%%%%%%%%%%%%%%%%%%%%%%%
 where $|b>$ and $|\beta>$ are bosonic and fermionic coherent states given by 
 $|b>=e^{-\frac{1}{2}|b|^2}e^{b\hat{b}^{\dagger}}|0>$ and 
 $|\beta>=e^{-\frac{1}{2}\beta^*\beta}e^{\hat{\beta}^{\dagger}\beta}|0>$.
In the symmetric gauge, the supercoherent state is written  as
%%%%%%%%%%%%%%%%%%%%%%%%%%%%%%%%%%%%%%%%%%%%%%%%%%%%%%%%%%%%%%%%%%%%%%%%%
\begin{equation}
\psi_{{b},\beta}=\sqrt{\frac{2}{\pi}}e^{-\frac{1}{2}(|b|^2+\beta^*\beta)}
e^{\sqrt{2}(bz^*+\theta\beta)}e^{-|z|^2-\theta\theta^*}.
\end{equation}
%%%%%%%%%%%%%%%%%%%%%%%%%%%%%%%%%%%%%%%%%%%%%%%%%%%%%%%%%%%%%%%%%%%%%%%%%%
We define a super von Neumann basis as a subset of the supercoherent states, 
whose index takes  discrete values
%%%%%%%%%%%%%%%%%%%%%%%%%%%%%%%%%%%%%%%%%%%%%%%%%%%%%%%%%%%%%%%%%%%%%%%%%%%%%  
\begin{equation}
b_{mn}=\sqrt{\pi}(m+in),
\end{equation}
%%%%%%%%%%%%%%%%%%%%%%%%%%%%%%%%%%%%%%%%%%%%%%%%%%%%%%%%%%%%%%%%%%%%%%%%%%
where $m$ and $n$ take integers.
It is easily checked that the 
complete relation for the super von Neumann basis 
 exactly coincides with the ``complete relation'' in the LLL (\ref{completeLLL}).
Thus, the super von Neumann basis spans the Hilbert space in the LLL.

\end{document}